\documentclass[12pt]{spieman}

\usepackage{amsmath,amsfonts,amssymb}

\usepackage{setspace}
\usepackage{tocloft}
\usepackage[]{graphicx, epstopdf}
\usepackage{indentfirst}
\usepackage{afterpage}
\usepackage{rotating}

\usepackage{hyperref}

\usepackage{mathrsfs}

\usepackage{wrapfig}
\usepackage{textcomp}

\usepackage{multirow}

\usepackage{xcolor}
\usepackage{ulem}

\newcommand{\tauwfs}{\tau_\mathrm{wfs}}

\newcommand{\Ji}{\mathrm{Ji}}

\newcommand{\karman}{K\'{a}rm\'{a}n }

\DeclareMathAlphabet{\mathpzc}{OT1}{pzc}{m}{it}

\newcommand{\argmin}{\operatornamewithlimits{argmin}}

\title{Ground-based adaptive optics coronagraphic performance under closed-loop predictive control}

\author[a]{Jared R. Males}
\author[a,b,c,d]{Olivier Guyon}
\affil[a]{Steward Observatory, University of Arizona, Tucson, 933 N Cherry Ave, Tucson, AZ 85721, USA}
\affil[b]{Astrobiology Center, National Institutes of Natural Sciences, 2-21-1 Osawa, Mitaka, Tokyo, JAPAN}
\affil[c]{College of Optical Science, University of Arizona, 1630 E University Blvd, Tucson, AZ 85719, USA}
\affil[d]{National Astronomical Observatory of Japan, Subaru Telescope, National Institutes of Natural Sciences, Hilo, HI 96720, USA}

\cftpagenumbersoff{figure}
\cftpagenumbersoff{table} 

\begin{document}
\maketitle

\begin{abstract}
The discovery of the exoplanet Proxima b highlights the potential for the coming generation of giant segmented mirror telescopes (GSMTs) to characterize terrestrial --- potentially habitable --- planets orbiting nearby stars with direct imaging.  This will require continued development and implementation of optimized adaptive optics systems feeding coronagraphs on the GSMTs.  Such development should proceed with an understanding of the fundamental limits imposed by atmospheric turbulence.  Here we seek to address this question with a semi-analytic framework for calculating the post-coronagraph contrast in a closed-loop AO system.  We do this starting with the temporal power spectra of the Fourier basis calculated assuming frozen flow turbulence, and then apply closed-loop transfer functions.  We include the benefits of a simple predictive controller, which we show could provide over a factor of 1400 gain in raw PSF contrast at 1 $\lambda/D$ on bright stars, and more than a factor of 30 gain on an I = 7.5 mag star such as Proxima.  More sophisticated predictive control can be expected to improve this even further.   Assuming a photon noise limited observing technique such as High Dispersion Coronagraphy, these gains in raw contrast will decrease integration times by the same large factors.  Predictive control of atmospheric turbulence should therefore be seen as one of the key technologies which will enable ground-based telescopes to characterize terrrestrial planets.
\end{abstract}

Keywords: Coronagraph, Adaptive Optics, Exoplanets, High-contrast Imaging

\section{Introduction}

The quest to characterize Earth-like planets was brought into sharp focus by the discovery of a terrestrial exoplanet orbiting the nearest star, Proxima Centauri \cite{2016Natur.536..437A}.  With a maximum projected separation of around 36 mas, Proxima b could in principle be characterized with current generation instruments \cite{2017A&A...599A..16L}, and will be readily studied with next generation giant segmented mirror telescopes (GSMTs) such as the 25 m Giant Magelllan Telescope (GMT), 30 m Thirty Meter Telescope (TMT), and 39 m Extremely Large Telescope (ELT).    Indeed, such observations are recognized as one of the key science goals of the GSMTs \cite{2012SPIE.8447E..1XG, 2014SPIE.9148E..20M}, and Proxima b will not be the only target.  Results from the {\it Kepler} mission \cite{2010Sci...327..977B} have made it clear that small planets occur frequently around low-mass stars \cite{2015ApJ...807...45D, 2015ApJ...798..112M}.  In addition to Proxima b, the recent discoveries of Ross 128 b\cite{2017arXiv171106177B}, LHS 1140 b\cite{2017Natur.544..333D}, and the seven terrestrial planets orbiting TRAPPIST-1\cite{2017Natur.542..456G} likewise point to a high frequency of terrestrial planets in low-mass star systems near the Earth.  

%\footnote{We follow the 2010 Decadal Report \cite{NAP12951} in using GSMT, rather than the currently common ELT as a general term, to avoid confusion with ESO's Extremely Large Telescope.} 

While some nearby planets, such as those orbiting TRAPPIST-1 and LHS 1140 b, will be amenable to atmospheric characterization with the transit spectroscopy technique, the low {\it a priori} transit probability means that the majority of nearby planets will have to be characterized with resolved direct imaging.  Low-mass stars present unique challenges and opportunities.  With lower stellar luminosity, planets with the same equilibrium temperature occur closer to the star.  This results in a significant improvement in contrast between the planet and the star, but also results in needing large diameter telescopes to resolve planets at such small separations \cite{2014SPIE.9148E..20M}.  This places such observations in the realm of ground-based telescopes equipped with adaptive optics (AO) and coronagraphs.  

The analysis presented here is motivated by the question: what are the fundamental limits of ground-based high-contrast coronagraphic imaging of terrestrial exoplanets?  Here we attack one part of this complicated issue: finding the fundamental limit imposed on coronagraphic contrast by residual atmospheric turbulence behind a closed-loop AO control system.  

Much previous work has been done modeling AO performance in the spatial-frequency domain  \cite{1998SPIE.3353.1038R, 2005JOSAA..22..310E,2006JOSAA..23..382J,2017JOSAA..34.1877C}.  This technique treats the AO system as a spatial filter, casting time and temporal-frequency domain processes (such as measurement noise and control filtering) into the the spatial domain as filters, and analyzing the effect of the combined AO system on the input turbulence power spectral density (PSD).

The previous analytic treatment of AO performance most closely related to our treatment here is Guyon (2005)\cite{2005ApJ...629..592G}.   Using the correspondence between Fourier modes and speckle amplitude, coupled with straightforward scaling from the input PSD, a derivation of the post-coronagraph contrast and its dependence on various AO system parameters was developed.  The resulting scaling laws are readily applicable to system design and analysis, and have proven to be useful in placing results from RV and transit surveys in context\cite{2013A&A...551A..99C}.

The goal of this study is to present a way to analyze the achievable contrast in a closed-loop AO system, including both current generation telescopes and future giant segmented mirror telescopes (GSMTs).  To do this we develop a semi-analytic framework in the temporal frequency domain.  That is, we explicitly calculate temporal PSDs rather than using the spatial-filtering approach.  At least in the frozen-flow regime, these should produce nearly identical results.  Our exploration of this method was motivated by the fact that actual AO systems must be optimized and controlled in the time and temporal-frequency domains. 

The paper is organized as follows. In Section \ref{sec:assumptions} we state our entering assumptions.  In Section \ref{sec:notation} we lay out our notation and physical description of the problem, including the statistics of the Fourier basis. In Section \ref{sec:psf_contrast} we derive a coronagraph model and deal with a subtle detail depending on how residual variance is calculated. In Section \ref{sec:temporal_psds} we calculate the temporal PSDs of Fourier modes assuming von \karman turbulence in frozen flow.  In Section \ref{sec:controls} we review the standard model of AO control with optimal gains, and present a predictive controller based on the Linear Prediction formalism.  Then, in Section \ref{sec:calcs} we use the previous results to predict the post-coronagraph contrasts for 6.5 m and 25.4 m telescopes for a range of guide star magnitudes, and compare optimal integrator and linear predictive controllers.  We discuss our results in Section \ref{sec:discussion} and conclude in Section \ref{sec:conclusion}.

\section{Assumptions}
\label{sec:assumptions}

This analysis is concerned with AO in the ``extreme'' regime: high actuator counts ($\sim$2000 on a 6.5 m aperture, $\sim$21,000 on a 25.4 m), and high loop speeds ($\sim$2000 Hz or more).  Furthermore, we are mainly interested in the coronagraphic raw contrast, that is the ratio of the intensity of the PSF at some separation from the star, to the intensity at the peak of the star's image or the point spread function (PSF).  Such extreme-AO (ExAO) systems are typically constructed for the purpose of high-contrast imaging of exoplanets and other narrow-angle circumstellar targets.

We do not seek to describe any particular ExAO system, nor address the details of design.  Rather, our goal is to assess what can be achieved with an ideal system limited only by the reality of atmospheric turbulence.    As such, or main assumptions are as follows:
\begin{enumerate}
\item We consider only on-axis guide stars, and neglect anisoplanatism.  This is of minor concern at the small angular separations we are concerned with for exoplanet imaging.
\item Our analysis will be monochromatic.   We do not consider the effects of atmospheric dispersion or chromaticity of the air index of refraction or scintillation within the wavefront sensor (WFS) band.  We also assume that the WFS and science wavelengths are the same, ignoring differential chromatic effects. 
\item We ignore WFS aliasing.  This is an approximation, but with justification: we analyze a pyramid WFS (PyWFS), which is less susceptible to aliasing than the Shack-Hartmann WFS \cite{2005MNRAS.357L..26V}.  Furthermore, a spatial filter \cite{2004JOSAA..21..810P} could be incorporated in a PyWFS essentially eliminating aliasing from concern.
\item We assume perfect knowledge of the AO system components.  This means we neglect errors in calibration (e.g. WFS gain) and assume that we know the spatial and temporal transfer functions of all components.
\item We will ignore all non-Kolmogorov wavefront error sources.  This means we do not consider static and non-common path errors within the coronagraph, nor do we analyze telescope error sources (e.g. vibrations).  We also consider only free atmosphere turbulence, that is we do not consider dome seeing. Such error sources are ultimately important in real systems, however free atmosphere turbulence is the dominant term which must be dealt with first. 
\item We will not consider changes in turbulence, e.g. changes in the wind velocity or changes in seeing.  
\item We will not consider details of coronagraphic architecture.  However, we do show that our idealized model is representative of real-world coronagraphs.
\end{enumerate}

Thus, what follows can be taken as goal setting:  we desire to know how well a coronagraph could perform if limited only by atmospheric turbulence.

\begin{table}[t]
\footnotesize
\centering
\caption{Summary of Notation \label{tab:notation}}
\begin{tabular}{ccl}
    Symbol                 & Units    & Parameter                    \\
\hline
\hline
$\vec{q}$                  &    m            & Position vector in the pupil plane \\
$u,v$                      &    m            & Position coordinates in the pupil plane \\
$\hat{\mathpzc{u}},\hat{\mathpzc{v}}$ &     & Coordinate unit vectors in the pupil plane\\
$\vec{k}$                  &   m$^{-1}$      & Spatial frequency vector in the pupil plane \\
$\Delta k$                 & m$^{-1}$        & Discrete spatial frequency sampling\\
$k_u,k_v$                  &   m$^{-1}$      & Spatial frequency components in the pupil plane \\
$m,n$                      &                 & Integer indices of discrete spatial frequencies \\
$D$                        & m               & Telescope diameter\\
$\mathscr{A}(\vec{q})$     &                 & Aperture function \\
$\lambda$                  &  m              & Wavelength of observation and wavefront sensing \\
$\mathcal{P}$              & rad$^2$ m$^2$   & Spatial power spectrum of the wavefront\\
$\lambda_0$                & m               & Reference wavelength for turbulence parameters \\
$r_0$                      & m               & Fried parameter for atmospheric turbulence strength\\
%$L_0$                      & m             & Outerscale of turbulence \\
%$k_0$                      & m$^{-1}$        & Spatial frequency of the outerscale \\
$Q $             &              & Denotes the Fourier transform of a mode\\
$z$                        & m            & Height above the observatory\\
$C_n^2(z_i)$               &              & Normalized turbulence strength for layer $i$\\
$M_{mn}$                   &              & Denotes a Fourier mode \\
$h_{mn}$                   & m            & Amplitude of a mode describing the wavefront phase \\
$a_{mn}$                   &              & Amplitude of a mode describing the wavefront amplitude \\
$\Phi(\vec{q},t)$          & rad          & The phase of the wavefront \\
$A(\vec{q},t)$             &              & The amplitude of the wavefront \\
$\sigma_{mn}^2$            & rad$^2$      & The variance of the amplitude of a mode \\
$\mathpzc{V}$              & m/s          & Wind speed\\
$\Theta$                   &  rad         & Wind direction\\
$\tau$                     & s            & Denotes a time delay\\
$\vec{r}$                  & $\lambda/D$  & Position vector in the image plane\\
$I$                        & photons/s/$(\lambda/D)^2$ & Intensity in the image plane \\
PSF                        & photons/s/$(\lambda/D)^2$ & Point-spread function\\
$S$                        &                             & Strehl ratio\\
$C$                        &                             & Contrast ratio\\
$f$                        & Hz           & Temporal frequency\\
$\mathcal{T}_{mn}$              & rad$^2$ / Hz & Temporal power spectrum of a single mode\\
$H(s)$                     &               & Denotes a transfer function \\

\hline
\end{tabular}
\end{table}

\section{The Incoming Wavefront}
\label{sec:notation}
We begin by introducing our notation and physical description of the problem.  We summarize our notation in Table \ref{tab:notation}.

\subsection{Coordinate System}

We will describe the incoming wavefront at the entrance pupil of the telescope.  The position vector $\vec{q}$ in this plane is defined by the unit vectors $(\hat{\mathpzc{u}},\hat{\mathpzc{v}})$, such that
\begin{equation}
\textstyle\vec{q} = u \hat{\mathpzc{u}} + v \hat{\mathpzc{v}} \\
\label{eqn:q_def}
\end{equation}
We will restrict our analysis to the unobstructed circular aperture of diameter $D$ defined by
\begin{equation}
\mathscr{A}(\vec{q}) = 
\begin{cases}
   \frac{4}{\pi D^2} , & \text{if } q \le \frac{D}{2} \\
   0, & \text{if } q > \frac{D}{2}.
\end{cases}
\label{eqn:unob_ap}
\end{equation}

Spatial-frequency is a vector quantity: 
\begin{equation}
\vec{k} = k_u \hat{\mathpzc{u}} + k_v \hat{\mathpzc{v}}. 
\end{equation}
Wavefront control is inherently a discrete problem, and the aperture defines the discrete spatial-frequency sampling:
\begin{equation}
\Delta k = \frac{1}{D}.
\end{equation}
Hence, we will often discretize spatial-frequency as
\begin{equation}
\textstyle\vec{k}_{mn} = \textstyle\frac{m}{D} \hat{\mathpzc{u}} + \textstyle\frac{n}{D} \hat{\mathpzc{v}}
\label{eqn:kq_def}
\end{equation}
where $m$ and $n$ are integer indices.

\subsection{Spatial Power Spectral Densities}

The spatial power spectral density (PSD) of the phase at the observation wavelength $\lambda$ in the von K\'{a}rm\'{a}n model \cite{1998aoat.book.....H} is
\begin{equation}
\mathcal{P}_{\mathrm{vK}}(\vec{k}, \lambda) = 0.0218 \left(\frac{\lambda_0}{\lambda}\right)^2\frac{1}{r_0^{5/3}} \frac{1}{(k^2 + k_0^2)^{11/6}} \mbox{ [rad$^2 / $m$^{-2}$]}
\label{eqn:psd_vK}
\end{equation}
where $r_0$ is Fried's parameter \cite{1965JOSA...55.1427F} and $\lambda_0$ is the wavelength at which $r_0$ is reported.  The outer scale, $L_0$, is included in the parameter 
\begin{equation}
k_0 = \frac{1}{L_0}.
\end{equation}
% The constant of proportionality is 
% \begin{equation}
% A_\mathcal{P} = \frac{\Gamma^2(11/6) \sin(5\pi/6)}{\pi^{11/3}}\left[ \frac{24}{5}\Gamma(6/5) \right]^{5/6} \approx 0.0218
% \end{equation}
% Equation \ref{eqn:psd_vK} reverts to the Kolmogorov PSD \cite{1976JOSA...66..207N} for large $L_0$, so we treat the von K\'{a}rm\'{a}n spectrum as the general case.

The mean value over the aperture, or the piston term, has no effect on our problem.  However, the amplitude of the piston mode is non-zero in the above PSD, so we should subtract the power in piston from the PSD.  The spatial PSD of a mode on the aperture is given by \cite{1993JOSAA..10..957R, 1995JOSAA..12.1559C}
\begin{equation}
\mathcal{P}_{\mathrm{mode}}(\vec{k}) = \mathcal{P}_{\mbox{o}}(\vec{k}) \left|  {Q}_{\mathrm{mode}}(\vec{k})   \right|^2
\label{eqn:psd_spatial}
\end{equation}
where $\mathcal{P}_{\mbox{o}}(\vec{k})$ is the input PSD ignoring the aperture, and ${Q}_{\mathrm{mode}}$ denotes the Fourier transform of the mode on the aperture.  As a Zernike polynomial, piston is simply $Z_{\mathrm{piston}} = 1$, and its Fourier transform on $\mathscr{A}(\vec{q})$ is \cite{1976JOSA...66..207N}
\begin{equation}
Q_{\mathrm{piston}} = 2 \frac{\mathrm{J}_1(\pi D k )}{\pi D k}\\
\end{equation}
So the piston subtracted phase PSD is
\begin{equation}
\mathcal{P}_{\mathrm{vK},\mathrm{sub}}(\vec{k}, \lambda) = \mathcal{P}_{\mathrm{vK}}(\vec{k}, \lambda)\left[1 - (2\Ji(\pi D k))^2  \right]
\label{eqn:psd_subtracted}
\end{equation}
where the Jinc function is
\begin{equation}
\Ji(x) = \frac{\mathrm{J}_1(x)}{x}
\end{equation}
and $J_1$ is the cylindrical Bessel function of the first kind. 

We model atmospheric turbulence as occurring in discrete layers.  To account for the Fresnel propagation between these layers we use the  functions\cite{2005ApJ...629..592G} 
\begin{eqnarray}
X(k,\lambda) &=& \sum_i C_n^2(z_i) \cos^2(\pi z_i k^2 \lambda) \nonumber \\
Y(k,\lambda) &=& \sum_i C_n^2(z_i) \sin^2(\pi z_i k^2 \lambda) 
\label{eqn:XandY}
\end{eqnarray}
where $C_n^2(z)$ is the normalized turbulence strength profile as a function of altitude $z$.  With these corrections, we have the PSD of the phase  
\begin{equation}
\mathcal{P}_\Phi(\vec{k}) =  \mathcal{P}_{vK,\mathrm{sub}}(\vec{k}, \lambda) X(k,\lambda)
\label{eqn:psd_phase_fresnel}
\end{equation}
and the PSD of the amplitude
\begin{equation}
\mathcal{P}_{A}(\vec{k}) = \mathcal{P}_{vK,\mathrm{sub}}(\vec{k}, \lambda) Y(k,\lambda)
\label{eqn:psd_amplitude_fresnel}
\end{equation}
at the aperture of the telescope.

%\subsection{Sampling the Wavefront}

\subsection{Representing the Wavefront}

\subsubsection{The Fourier Basis}
\label{sec:basic_fourier}
Here we develop a real-valued form of the Fourier basis. Consider the complex exponential form of the basis \cite{2007JOSAA..24.2645P}:
\begin{equation}
M_{mn} = \mathpzc{h} e^{i (2\pi \vec{k}_{mn} \cdot \vec{q})}
\end{equation}
where we have included a complex-valued coefficient $\mathpzc{h}$ to illustrate the form of this function.  Decomposing this into sines and cosines we have
\begin{equation}
M_{mn} = \mbox{Re}\{\mathpzc{h}\}( \cos(2\pi \vec{k}_{mn} \cdot \vec{q}) + i \sin(2\pi \vec{k}_{mn} \cdot \vec{q})) + \mbox{Im}\{\mathpzc{h}\}(i\cos(2\pi \vec{k}_{mn} \cdot \vec{q}) - \sin(2\pi \vec{k}_{mn} \cdot \vec{q}))
\end{equation}
where Re$\{\cdot\}$ and Im$\{\cdot\}$ denote the real and imaginary part. This suggests a real-valued Fourier basis defined as: 
\begin{equation}
M_{mn}^p(\vec{q}) = \cos(2\pi \vec{k}_{mn} \cdot \vec{q}) + p  \sin(2\pi \vec{k}_{mn} \cdot \vec{q})
\label{eqn:modified_fourier}
\end{equation}
where $p=\pm 1$ (in the superscript we will use $+$ and $-$).  These modes are neither odd nor even, and are normalized
\begin{equation}
\textstyle\int \mathscr{A}(\vec{q}) \left(M_{mn}^p(\vec{q})\right)^2 d\vec{q} = 1, \\
\end{equation}
though not orthogonal, on the aperture.  For completeness we show how the normalization is derived in Appendix \ref{app:norm}. The Fourier transforms on the aperture are (see Appendix \ref{app:ft})
\begin{equation}
{Q}_{mn}^p (\vec{k}) =  (1+i^p)\Ji( \pi D k_{mn}^+) + (1-i^p)\Ji( \pi D k_{mn}^-). 
\label{eqn:modf_ft}
\end{equation}
where, for notational simplicity, we have defined
\begin{align}
\begin{split}
k_{mn}^+ &= \sqrt{ \left(k_u + \textstyle\frac{m}{D} \right)^2 + \left(k_v + \textstyle\frac{n}{D} \right)^2 }\\
k_{mn}^- &= \sqrt{ \left(k_u - \textstyle\frac{m}{D} \right)^2 + \left(k_v - \textstyle\frac{n}{D} \right)^2 } 
\end{split}
\label{eqn:k_projected}
\end{align}

We have chosen this form of the Fourier basis primarily because it yields the useful result 
\begin{equation}
\left|{Q}_{mn}^+ (\vec{k})\right|^2 = \left|{Q}_{mn}^- (\vec{k})\right|^2
\end{equation}
This means that each spatial frequency can be described by a single spatial PSD, and, as we will show, a single one-sided (positive frequencies only) temporal PSD.

\subsubsection{The Wavefront in the Fourier Basis}
We denote the wavefront phase as $\Phi(\vec{q},t)$, which has units of radians. The coefficient of a mode is calculated as
\begin{equation}
h_{mn}^p(t) = \frac{\lambda}{2\pi} \int \mathscr{A}(\vec{q}) M_{mn}^p (\vec{q}) \Phi(\vec{q},t) d\vec{q}.
\end{equation}
These coefficients are real-valued.  Poyneer and Veran \cite{2005JOSAA..22.1515P} showed that the Fourier basis can be used for analysis and synthesis, even though it is not orthogonal on an aperture. Since our slightly modified basis is simply a linear combination of the cosine and sine, their results apply to it as well.  This means that we can write
\begin{equation}
\Phi(\vec{q},t) = \frac{2\pi}{\lambda} \sum\limits_{mn} \left[ h_{mn}^{+}(t) M_{mn}^{+}(\vec{q}) + h_{mn}^{-}(t) M_{mn}^{-}(\vec{q})\right]
\label{eqn:phi_expansion}
\end{equation}

Similarly, we represent the wavefront amplitude aberrations as
\begin{equation}
A(\vec{q},t) = 1 + \sum\limits_{mn}\left[ a_{mn}^{+}(t) M_{mn}^{+}(\vec{q}) + a_{mn}^{-}(t) M_{mn}^{-}(\vec{q})\right]
\end{equation}

% We will use this form of the Fourier basis for the remainder of our analysis.   Using it we need only one spatial PSD, and, later, a single one-sided temporal PSD for each spatial frequency, saving computation time.  One will also find that the uniform and constant normalization makes some of the coming algebra less cumbersome.  Of course, this analysis could be carried out using the conventional sine and cosine basis of Equation (\ref{eqn:basic_fourier}), or the complex exponential form, and it would yield the same results in the end. 

\subsection{Statistics of the Fourier Basis}

Following the recipe of Noll\cite{1976JOSA...66..207N}, we can write the covariance of any two modes over the aperture as
\begin{equation}
\left< h_{mn}^{p\ast} h_{m'n'}^{p'} \right>  =  \int_{0}^{\infty} \int_{0}^{2\pi} {Q}_{mn}^{p\ast}{Q}_{m'n'}^{p'} \mathcal{P}(k)  d\theta k dk.
\label{eqn:covar_gen}
\end{equation}
We now make use of this expression to analyze the statistics of the Fourier basis under von \karman turbulence.

\subsubsection{Ignoring the Aperture}
\label{seg:ignore_aperture}
In Equation (\ref{eqn:covar_gen}) the Fourier transform $Q$ describes how the PSD is sampled by the modes. If we were to ignore the spatial effect of the aperture, we would have
\begin{equation}
Q_{mn} \sim \delta(|\vec{k} - \vec{k}_{mn}|)\Delta k
\end{equation}
where $\delta(x)$ is the Dirac delta distribution, which is the Fourier transform of the sine and cosine on an infinite domain.  With this approximation, we have the variance of a mode  

\begin{equation}
\sigma^2_{mn} = \mathcal{P}_{\Phi}(\vec{k}_{mn}, \lambda) (\Delta k)^2 = \frac{\mathcal{P}_{\Phi}(\vec{k}_{mn}, \lambda)}{D^2} \mbox{ [radians]}.
\label{eqn:var_naive}
\end{equation}

\subsubsection{Modal Variance}
\label{sec:modal_variance}
Including the aperture, the variance of a mode in radians is given by
\begin{equation}
\sigma^2_{mn} = \left(\frac{2\pi}{\lambda}\right)^2\left< |h_{mn}^{p}|^2 \right> = \int_{0}^{\infty} \int_{0}^{2\pi} |{Q}_{mn}^{p}|^2 \mathcal{P}(k)  d\theta k dk.
\label{eqn:var_mode}
\end{equation}
Now since
\begin{equation}
\left|{Q}_{mn}^p \right|^2 = {Q}_{mn}^{p\ast}{Q}_{mn}^p = 2\left(\Ji^2(\pi D k_{mn}^+) + \Ji^2(\pi D k_{mn}^-)\right)
\end{equation}
independent of $p$, it follows that $\left< |h_{mn}^{+}|^2 \right> \: = \: \left< | h_{mn}^{-}| ^2 \right>$.   From now on we will drop the $p$ superscript except where it is explicitly needed.

\begin{figure}
\centering
\includegraphics[width=5in]{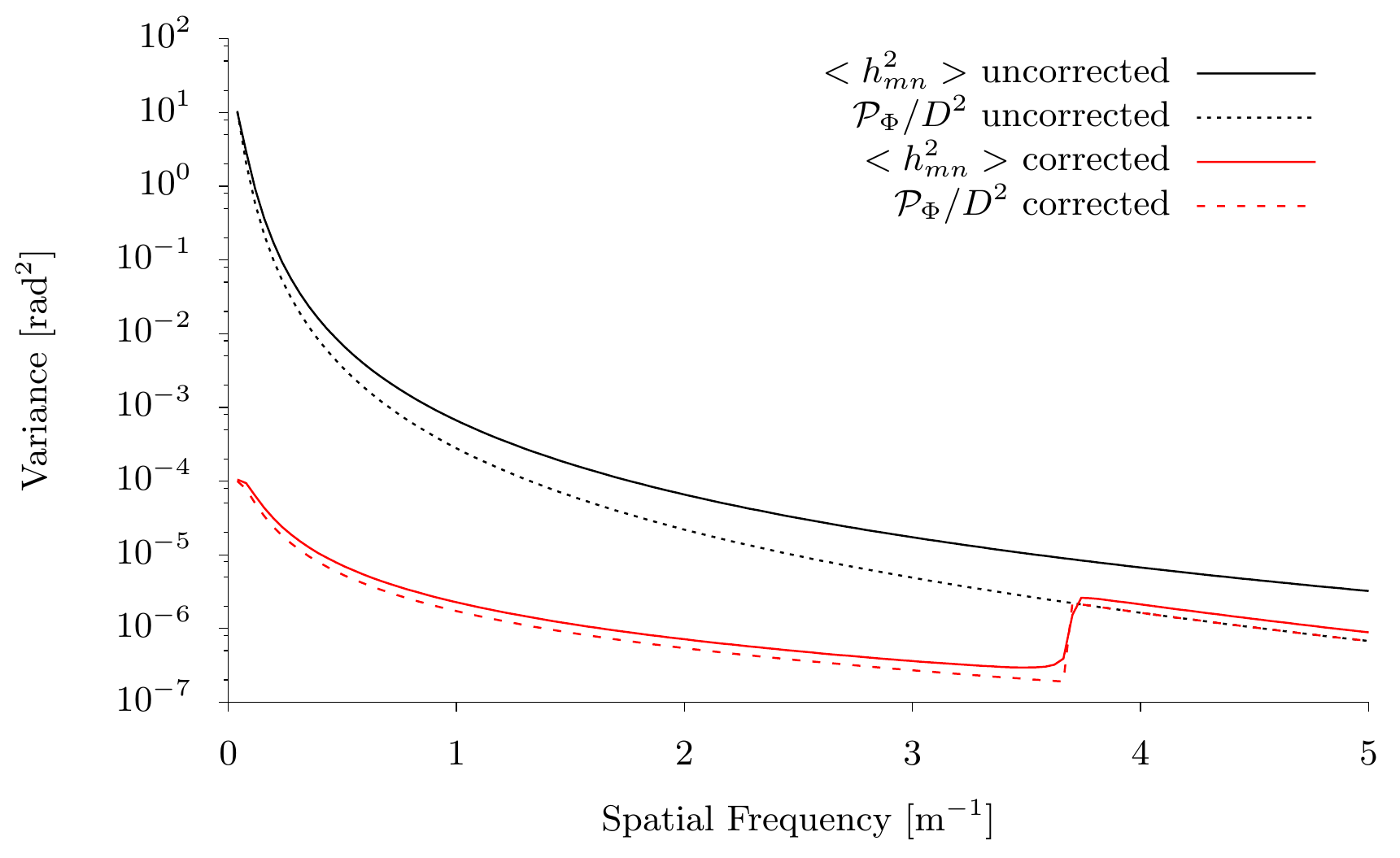}
\caption{ \label{fig:modevars} Modal variance vs. spatial frequency.  We compare the result of Equation (\ref{eqn:covar_gen}) (solid lines) with the simple estimate of $\mathcal{P}_\Phi/D^2$ (dashed lines).  The black curves are for uncorrected turbulence, and the red curves are for a simple treatment of AO correction.  The discrepancy between the solid and dashed curves is due to the convolution on the aperture, which is neglected in the simple estimate.}
\end{figure}

In Figure \ref{fig:modevars} we show the modal variance vs spatial frequency.  This was calculated using numerical integration in \verb double  precision with the GNU Scientific Library (GSL \cite{gsl_manual}) routines \\
\verb gsl_integration_qagiu  (for the radial integral) and \verb gsl_integration_qag  (for the azimuthal integral).  We found that an absolute tolerance of $10^{-10}$ and a relative tolerance of $10^{-4}$ gave good results.

Figure \ref{fig:modevars} also shows the naive result using Equation (\ref{eqn:var_naive}). The simple estimate based on the PSD is significantly below the true variance.  

One might guess that AO correction will reduce or eliminate this discrepancy.  A simple way to estimate the effect of AO control on these statistics is to assume frozen flow and apply a scaling for the time-delay error.  Following Guyon\cite{2005ApJ...629..592G} (as modified in Appendix \ref{app:guyon_contrast}) the corrected PSD is approximated by
\begin{equation}
\mathcal{P}_{\mathrm{cor}}( \vec{k}, \lambda) \approx 
\begin{cases}
\mathcal{P}_\Phi(\vec{k})\left(2 \pi \mathpzc{V} k\right)^2 \tau_{tl}^2, & k \le \frac{D}{2d}\\
\mathcal{P}_\Phi(\vec{k}), & k > \frac{D}{2d}
\end{cases}
\label{eqn:simple_est}
\end{equation}
where $\mathpzc{V}$ is the wind speed, $\tau_{tl}$ is the total latency in the control system, and $d$ is the actuator spacing projected on the aperture.  We next recalculated the modal variances with the corrected PSD, using $\mathpzc{V} = 10$ m/s, $\tau_{tl} = 1.25$ ms, and $d = 13.5$ cm on a $D = 6.5$ m aperture.

The variance with and without the aperture is also shown in Figure \ref{fig:modevars}.  Though the discrepancy is reduced, the simple estimate based on the PSD is still below the true variance. This means that the variance of a Fourier mode is not simply proportional to the PSD when it is sampled by the aperture.  

The discrepancy is due to the aperture: there is a convolution inherent in Equation (\ref{eqn:covar_gen}) not addressed by Equation (\ref{eqn:var_naive}).  We show how to address this, and discuss its implications for various coronagraph models, in Section \ref{sec:psf_contrast} below.

\subsubsection{Correlation}
%\begin{equation}
%\left<h_{mn}^+ h_{m'n'}^+ \right> + \left<h_{mn}^- h_{m'n'}^-\right> = 2\mathcal{R}\left\{ \left< h_{mn}^+ h_{m'n'}^+ \right> \right\} = 2\mathcal{R}\left\{ \left< h_{mn}^- h_{m'n'}^- \right> \right\}
%\end{equation}

%\begin{equation}
%\left<h_{mn}^+ h_{m'n'}^- \right> + \left<h_{mn}^- h_{m'n'}^+\right> = 2\mathcal{R}\left\{ \left< h_{mn}^+ h_{m'n'}^- \right> \right\} = 2\mathcal{R}\left\{ \left< h_{mn}^- h_{m'n'}^+ \right> \right\}
%\end{equation}

\begin{figure}
\centering
%\scalebox{0.8}{\input{figures/covar_normed_uncorrected}}
\includegraphics[width=7in]{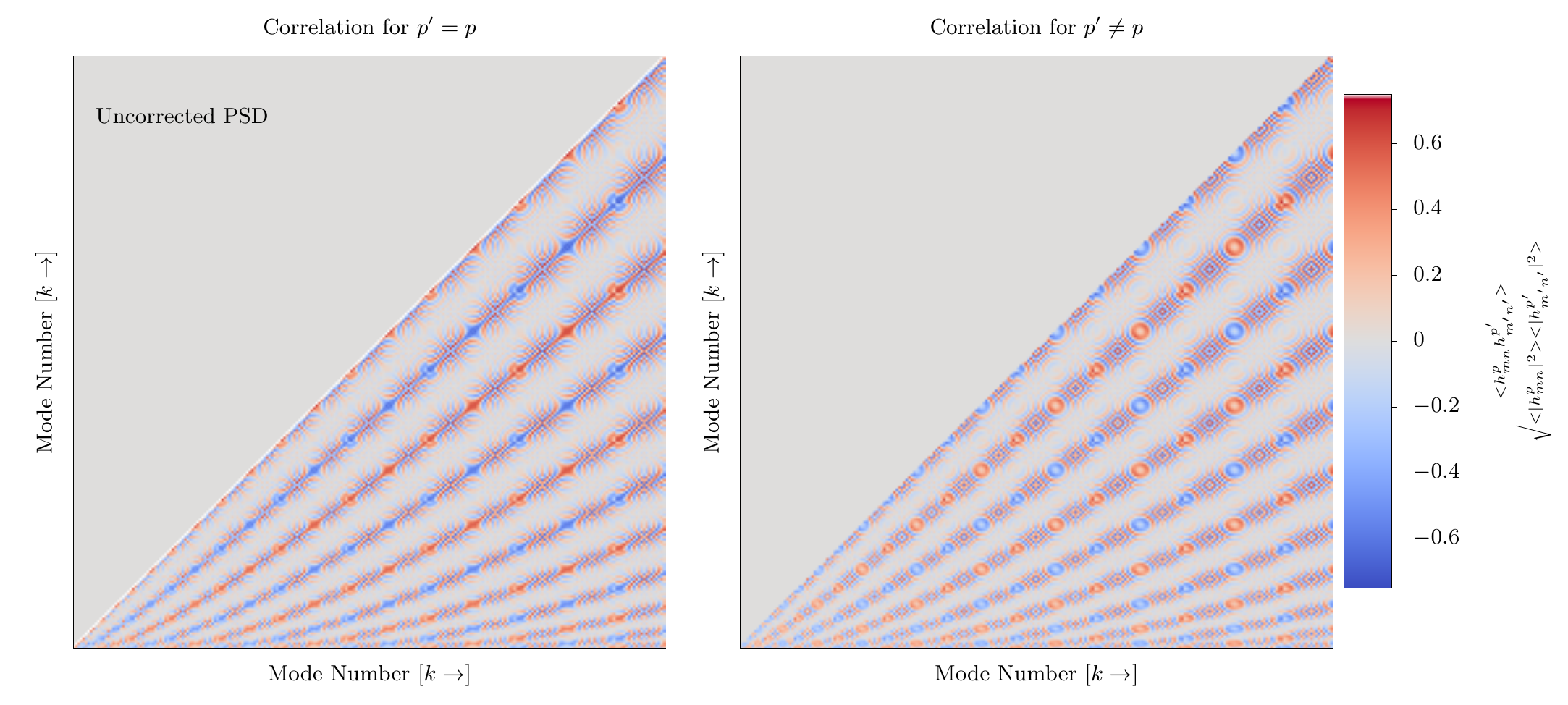}
\caption{Correlations between the Fourier modes in uncorrected von \karman turbulence.  At left we show modes with $p' = p$, and at right modes with $p' \ne p$.  Each covariance point is normalized to the associated diagonal in the left hand plot, which is the variance of each mode.  In uncorrected turbulence, the correlations are high relative to the modal variances.
\label{fig:covar_normed}}
\end{figure}

\begin{figure}
\centering
%\scalebox{0.8}{\input{figures/covar_normed_corrected}}
\includegraphics[width=7in]{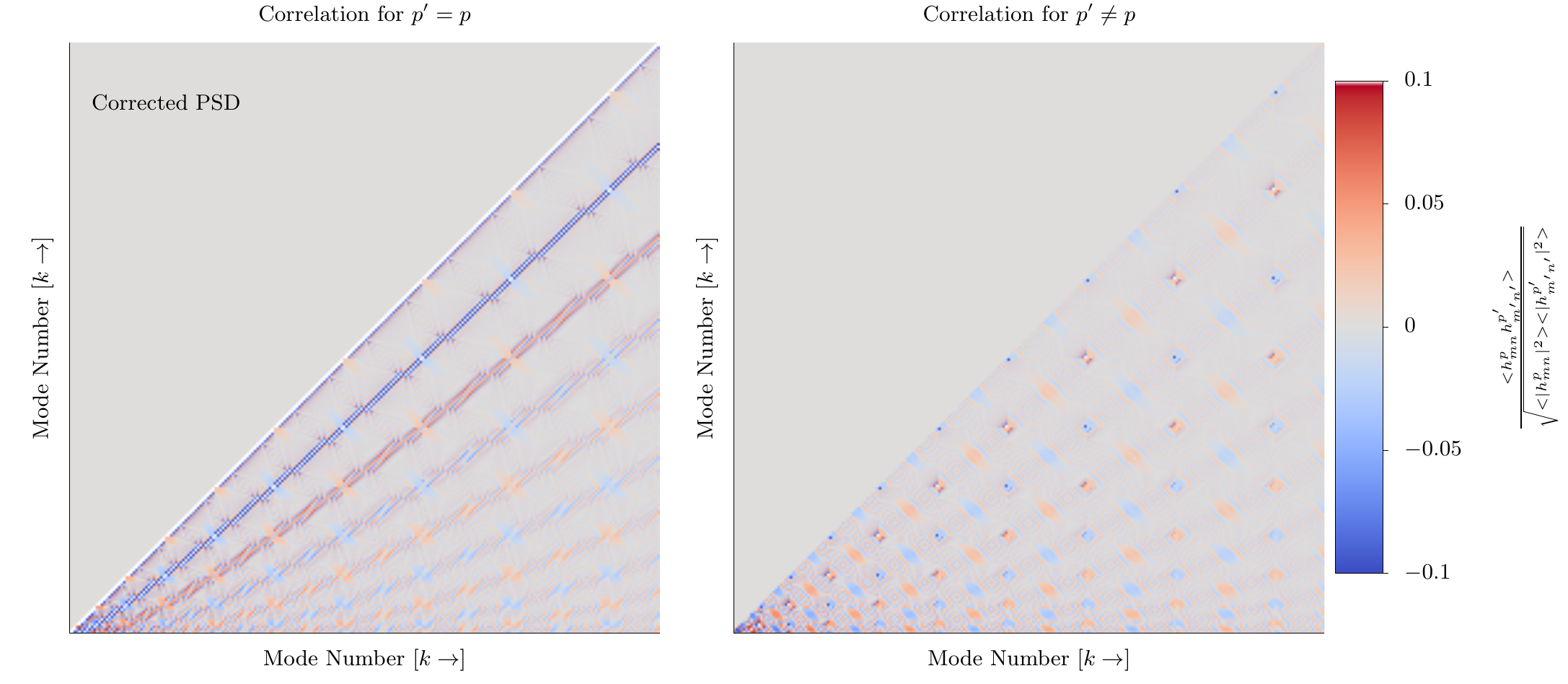}
\caption{Same as Figure \ref{fig:covar_normed}, but for AO corrected von \karman turbulence.  After AO correction removes low-spatial-frequency power, the modal correlations are much smaller but not zero. 
\label{fig:covar_normed_corrected}}
\end{figure}

We next calculated the covariance of pairs of Fourier modes according to Equation (\ref{eqn:covar_gen}).  The results are shown in Figure
\ref{fig:covar_normed} for uncorrected von \karman turbulence, and in Figure \ref{fig:covar_normed_corrected} for the corrected case.  In the figure we show the covariances normalized by the modal variances, that is the correlation.    There is significant correlation between various modes in the uncorrected case, but an AO system acts to suppress this.   In the following section we show that this correlation between Fourier modes can be accounted for by convolving the PSF with the PSD.

\section{Post-Coronagraph Contrast}
\label{sec:psf_contrast}
We next derive a PSD-based model of a coronagraph fed by an AO system in the long exposure limit, and assess its validity.  

\subsection{Coronagraph Model}
The intensity in an image plane is given by
\begin{equation}
I(\vec{r},t) = \left| \mathcal{F} \left\{  \mathscr{A}(\vec{q}) A(\vec{q},t) e^{i\Phi(\vec{q},t)}   \right\} \right|^2
\label{eqn:irrad_act}
\end{equation}
The operator $\mathcal{F}\left\{ \cdot \right\}$ describes the propagation of the wavefront from pupil to final image plane.  From here on we assume this is the Fourier transform defined as in Appendix \ref{app:ft}, though it could be a series of transforms corresponding to various coronagraph components\cite{2017MNRAS.467L.105H}.  We define the system PSF as the result of the above with no aberrations, that is $h_{mn}=0$ and  $a_{mn}=0$ for all $m,n$, which is
\begin{equation}
\mbox{PSF}(\vec{r}) = \left| \mathcal{F} \left\{  \mathscr{A}(\vec{q})   \right\} \right|^2
\label{eqn:psfdef}
\end{equation}

In the case of the circular unobstructed aperture (Equation \ref{eqn:unob_ap}) this is the Airy pattern 
\begin{equation}
\mbox{PSF}(\vec{r}) = \left( 2 \Ji( \pi D r /\lambda)  \right)^2
\label{eqn:airypsf}
\end{equation}
where we have used the image plane coordinate $\vec{r} = \lambda \vec{k}$.  Telescope apertures are typically more complicated, consisting of secondary mirror obscuration and support structures, possibly segments, and can be non-circular.  Furthermore, it could be position dependent due to the coronagraph.  The result will be more difficult to analyze, usually requiring numerical calculation.  

%In this work we use the simple unobstructed circular aperture for ease of analysis, but all the results apply to more complicated apertures.  

%We also restrict ourselves to monochromatic light for now.

Equation (\ref{eqn:irrad_act}) can not be evaluated in closed form in the general case.  However,  by assuming small aberrations and an ideal coronagraph we can use a 2nd order Taylor expansion \cite{2002ApJ...581L..59S,2003ApJ...596..702P} to approximate the post-coronagraph PSF as\cite{2005SPIE.5903..170M} \footnote{Macintosh et al. \cite{2005SPIE.5903..170M} only addressed phase, but the extension to amplitude is straightforward from their arguments.} 
\begin{equation}
I(\vec{r},t) \approx \left| \mathcal{F} \left\{ \mathscr{A}(\vec{q}) \textstyle\sum\limits_{mn}A_{mn}(\vec{q}, t)      \right\} \right|^2 + \left| \mathcal{F} \left\{ \mathscr{A}(\vec{q})  \Phi(\vec{q},t)      \right\} \right|^2
\label{eqn:perrin_psf}
\end{equation}

\begin{figure}[t]
\centering
\includegraphics[width=5in]{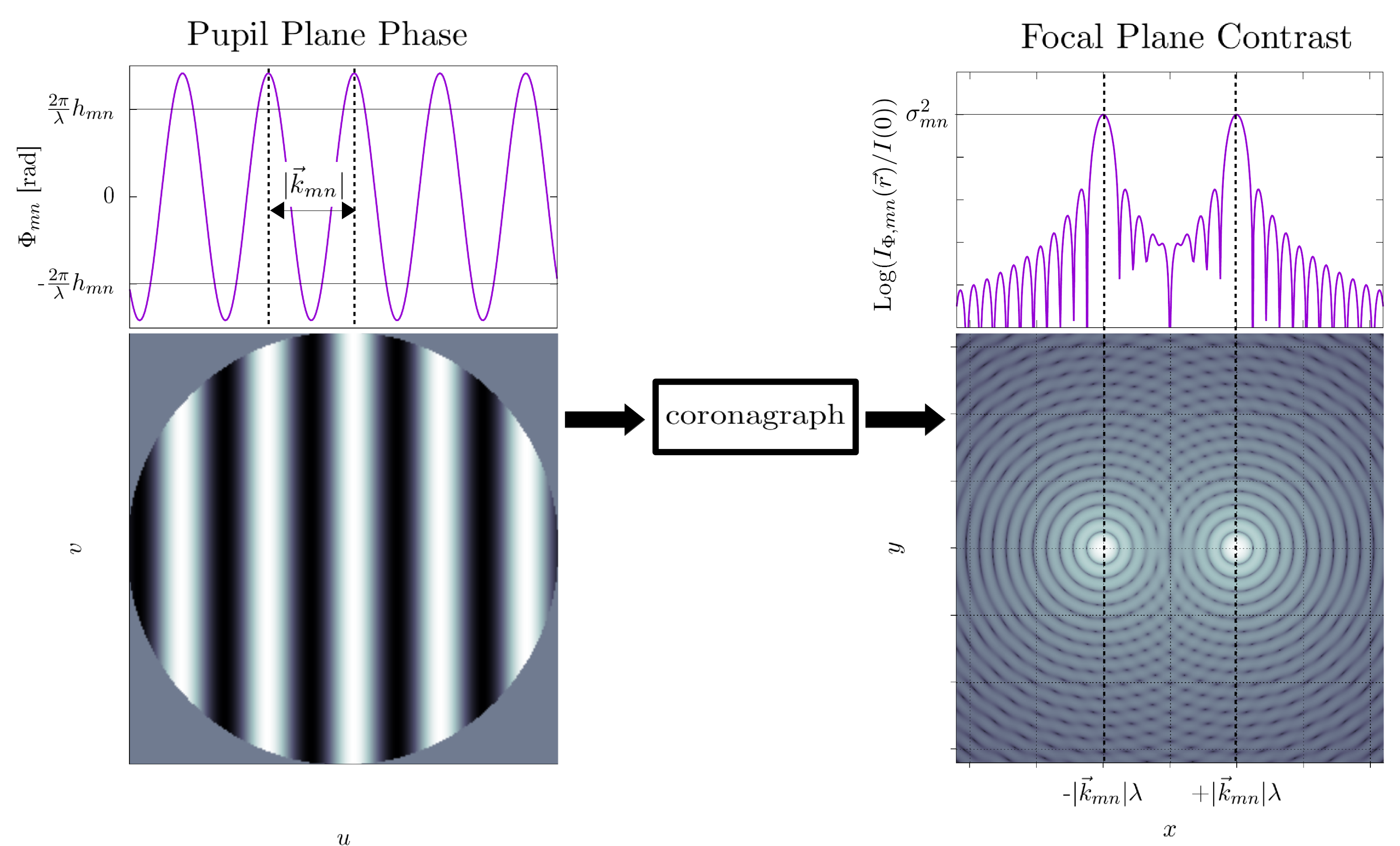}
\caption{ Illustration of the relationship between a phase aberration defined by a Fourier mode with spatial frequency $\vec{k}_{mn}$ and coefficient $h_{mn}$, to the long-exposure contrast of the resultant speckles in the post-coronagraph image plane.  This relationship does not fully define the contrast in the presence of many speckles.  \label{fig:illust} }
\end{figure}

Dealing with the phase component first, we have
\begin{equation}
I_\Phi(\vec{r},t)  \approx \left| \textstyle\sum\limits_{mn}  \mathcal{F} \left\{ \mathscr{A}(\vec{q})\Phi_{mn}(\vec{q},t)     \right\} \right|^2
\label{eqn:Iph}
\end{equation}
Evaluating a single term gives
\begin{equation}
\mathcal{F} \left\{ \mathscr{A}(\vec{q})\Phi_{mn}(\vec{q},t)     \right\}= \frac{2\pi}{\lambda} \left[ h_{mn}^{+}(t) {Q}_{mn}^{+}(\vec{k})  + h_{mn}^{-}(t)  {Q}_{mn}^{-}(\vec{k}) \right]
\end{equation}
Then taking the modulus squared and pulling out the terms with $mn = m'n'$ gives\footnote{We are suppressing the $t$ dependence of $h$.}
% \begin{equation}
% I_\Phi(\vec{r},t) = \left(\frac{2\pi}{\lambda}\right)^2 2 \sum_{mn} \left[  \left( h_{mn}^{+}(t)^2 + h_{mn}^{-}(t)^2 \right) \left( \Ji^2(\pi D k_{mn}^+) +\Ji^2(\pi D k_{mn}^-)\right) + \mbox{ cross-terms} \right]
% \end{equation}
\begin{eqnarray}
I_\Phi(\vec{r},t) &\approx&\left(\frac{2\pi}{\lambda}\right)^2 2 \sum_{mn} \left\{  \left[ (h_{mn}^{+})^2 + (h_{mn}^{-})^2 \right] \left[ \Ji^2(\pi D k_{mn}^+) +\Ji^2(\pi D k_{mn}^-)\right] \right. \nonumber \\
                  &+& 8 h_{mn}^+ h_{mn}^-  \Ji(\pi D k_{mn}^+) \Ji(\pi D k_{mn}^-)  \nonumber \\
                  + &\sum_{\substack{m'n' \\ mn \ne m'n'}}&\left\{ \left[h_{mn}^{+} h_{m'n'}^{+} +  h_{mn}^{-}h_{m'n'}^{-}\right] \left[\Ji(\pi D k_{mn}^+)\Ji(\pi D k_{m'n'}^+)+\Ji(\pi D k_{mn}^-)\Ji(\pi D k_{m'n'}^-) \right] \right. \label{eqn:I_sum_expanded} \\
                  && +  \left. \left. \left[h_{mn}^{+} h_{m'n'}^{-} + h_{mn}^{-} h_{m'n'}^{+}\right]\left[ \Ji(\pi D k_{mn}^+)\Ji(\pi D k_{m'n'}^-)+\Ji(\pi D k_{mn}^-)\Ji(\pi D k_{m'n'}^+)\right]  \right\} \right\}, \nonumber
\end{eqnarray}
which is the instantaneous intensity at a position $\vec{r}$ and time $t$ in the post-coronagraph image plane.  Now, we seek the long exposure, or expected, value of this expression.  Though this appears quite intractable, we see that this will require evaluating covariances of the form $<h_{mn}^p h_{m'n'}^{p'}>$.  So, to proceed we make the ansatz that we can temporarily ignore the aperture as in Section \ref{seg:ignore_aperture}.  The effect of this assumption is that all covariances are 0, which leads directly to
\begin{equation}
\left< I_{\Phi,mn}(\vec{r}) \right> = \frac{\mathcal{P}_{\Phi}(\vec{k}_{mn}, \lambda)}{D^2}  \left[ \mbox{PSF}(\vec{r}-\vec{k}_{mn}\lambda) +  \mbox{PSF}(\vec{r}+\vec{k}_{mn}\lambda)\right]
\label{eqn:contrast_var_mn}
\end{equation}
for a single term in the sum on the first line of Equation (\ref{eqn:I_sum_expanded}). This is the classic result of a Fourier aberration producing a pair of symmetric speckles \cite{1995PASP..107..386M}. We illustrate this in Figure \ref{fig:illust}.

% \begin{equation}
% \left< I_\Phi(\vec{r}) \right> \approx \left(\frac{2\pi}{\lambda}\right)^2 2 \sum_{mn} \left( \left<h_{mn}^{+\;2}\right> + \left<h_{mn}^{-\;2}\right> \right) \left( \Ji^2(\pi D k_{mn}^+) +\Ji^2(\pi D k_{mn}^-)\right) 
% \end{equation}
% 
% Next employing  we arrive at 
% \begin{equation}
% \left< I_{\Phi,mn}(\vec{r}) \right> \approx \left(\frac{\pi}{\lambda}\right)^2 2 \left( \left<h_{mn}^{+\;2}\right> + \left<h_{mn}^{-\;2}\right> \right) \left[ \mbox{PSF}(\vec{r}-\vec{k}_{mn}\lambda) +  \mbox{PSF}(\vec{r}+\vec{k}_{mn}\lambda)\right].
% \label{eqn:contrast_Phi_mn}
% \end{equation}
% With an extra factor of two due to normalization, this is the classic result of a Fourier aberration producing a pair of symmetric speckles \cite{1995PASP..107..386M}. 

Finally, we perform the sum over all spatial frequencies. The complete intensity due to $\Phi$ at $\vec{r}$ is
\begin{equation}
\left< I_\Phi(\vec{r}) \right> = \textstyle\sum\limits_{mn} \left< I_{\Phi,mn}(\vec{r}) \right>.
\label{eqn:contrast_Phi}
\end{equation}
This summation is an unnormalized convolution with the PSF, and can be thought of as adding photons from nearby speckles.  This step accounts for the effect of the aperture.  In the AO corrected regime, it increases the intensity by as much as 50\%, explaining the discrepancies in Figure \ref{fig:modevars}.  

Now we apply the lesson of Section \ref{sec:modal_variance} above.  If we instead use the variance calculated with Equation (\ref{eqn:var_mode}) , then we can skip the convolution and arrive directly at
\begin{equation}
\left< I_\Phi(\vec{r}) \right> =  \left(\frac{2\pi}{\lambda}\right)^2\left< |h_{mn}|^2 \right>  \left[ \mbox{PSF}(\vec{r}-\vec{k}_{mn}\lambda) +  \mbox{PSF}(\vec{r}+\vec{k}_{mn}\lambda)\right].
\label{eqn:contrast_h2}
\end{equation}
This is an assertion that the the complete evaluation of Equation \ref{eqn:I_sum_expanded} in the long exposure limit is equivalent to the modal variance calculation on the aperture with Equation \ref{eqn:var_mode}.  Acknowledging that we have not actually proven this, we demonstrate its validity with numerical experiments in the following section.

The amplitude component of the post-coronagraph long-exposure intensity, $\left< I_A(\vec{r}) \right>$, is evaluated in nearly identical fashion.  
%We don't repeat the above s
%\begin{equation}
%\left< I_{A,mn}(\vec{r}) \right>  = \left(\frac{1}{2}\right)^2 2\left( \left< a_{mn}^{+\;2}\right> + \left< a_{mn}^{-\;2}\right> \right)  \left[ \mbox{PSF}(\vec{r}-\vec{k}_{mn}\lambda) +  \mbox{PSF}(\vec{r}+\vec{k}_{mn}\lambda)\right].
%\label{eqn:contrast_A_mn}
%\end{equation}
%and the total intensity due to amplitude aberrations is
%\begin{equation}
%\left< I_A(\vec{r}) \right>  = \textstyle\sum\limits_{mn} \left< A_{mn}(\vec{r}) \right>.
%\label{eqn:contrast_A}
%\end{equation}

We define the ``raw PSF contrast'' as the ratio of the intensity at $\vec{r}$ to the peak intensity of a non-coronagraphic long-exposure image of the star, $\left<I_{\mathrm{NC}}(0)\right>$.  This is related to the PSF by the long-exposure Strehl ratio $\left< S \right>$, defined by the relationship
\begin{equation}
\left< I_{\mathrm{NC}}(0) \right> = \left< S \right> \: \mathrm{PSF}(0).
\end{equation}
The contrast is then
\begin{equation}
\left< C(\vec{r}) \right>  = \frac{ \left< I_\Phi(\vec{r}) \right>  + \left< I_A(\vec{r})\right> }{ \left< S \right> \: \mathrm{PSF}(0)}
\label{eqn:psf_contrast}
\end{equation}

\subsection{Numerical Verification}

\begin{figure}[t!]
\footnotesize
\centering
\includegraphics[width=6.5in]{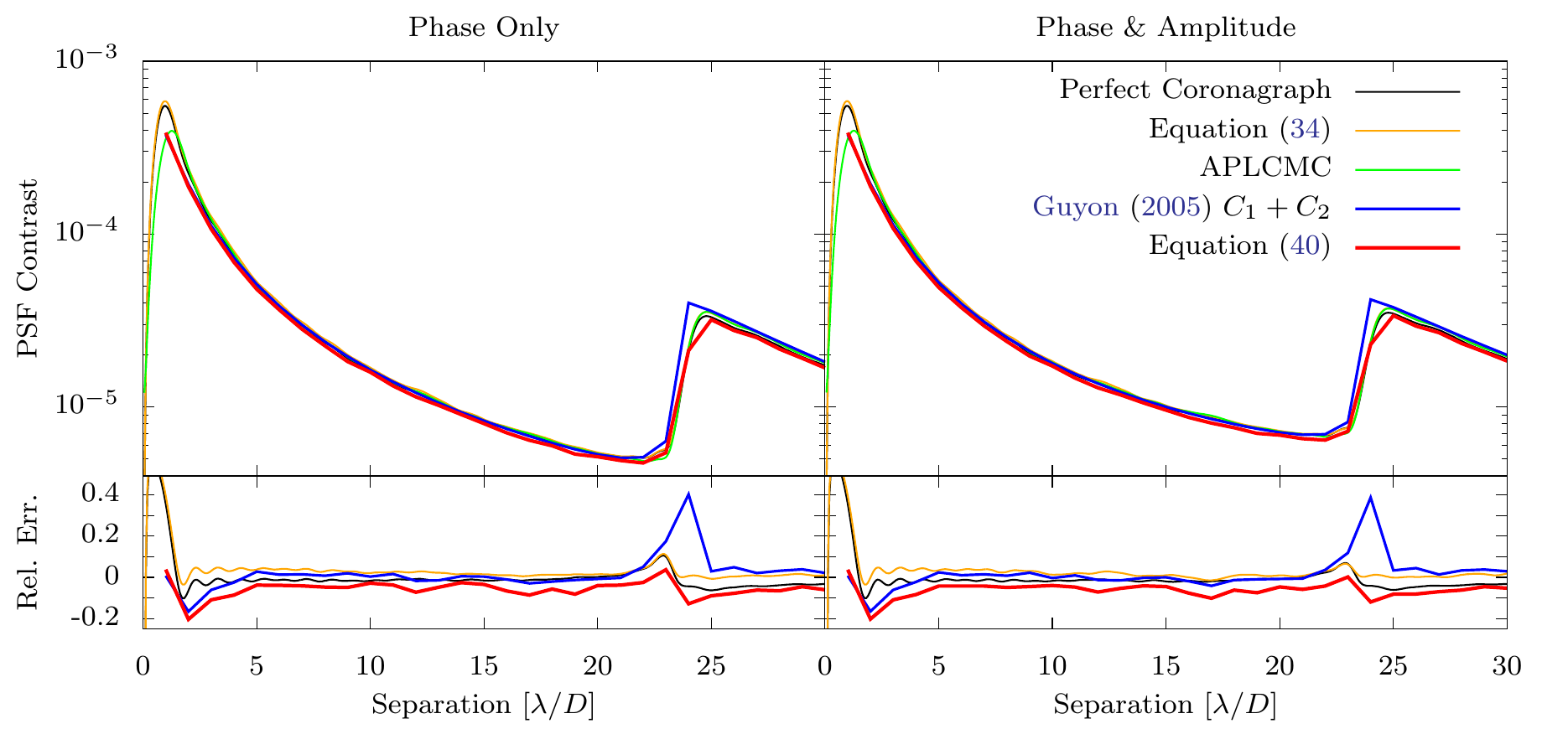}
\vspace{0.0in}
\caption{Numerical verification of coronagraph models.  Here we compare PSF contrasts for the perfect coronagraph \cite{2006A&A...447..397C}, the coronagraph defined by Equation (\ref{eqn:perrin_psf}), the APLCMC \cite{2014ApJ...780..171G}, the contrasts $C_1 + C_2$ from Guyon\cite{2005ApJ...629..592G}, and the variance of the Fourier modes as in Equation (\ref{eqn:contrast_h2}).  The left panel is for residual atmospheric phase aberrations only, and the right includes amplitude errors.  The bottom panel shows the fractional errors of the model coronagraph w.r.t. the APLCMC. In general, the analytic coronagraph models match the physically realizable APLCMC well.  This shows that such models are valid for analyzing the potential of ground-based AO-fed coronagraphs.
\label{fig:coronVerification}}
\end{figure}

We conducted a set of numerical experiments to verify our long exposure coronagraph model, compare it to other models, and to assess its relevance to real-world coronagraphy.  Five distinct model coronagraphs were considered, in each case both for phase aberrations only and with the amplitude errors included.  

The first comparison coronagraph model was the so-called ``perfect coronagraph'' \cite{2006A&A...447..397C}.  In this model the complex plane wave which minimizes the energy in the pupil is subtracted, and the result is propagated to the image plane.  It has been shown that analysis with the perfect coronagraph produces results which match the real-world Four Quadrant Phase Mask (4QPM) 
coronagraph \cite{2010JOSAA..27A.157S, 2000PASP..112.1479R}. 

The second coronagraph model consisted of propagating the complex wavefronts using Equation (\ref{eqn:perrin_psf}) directly, employing the FFT.  The phase and amplitude components were treated separately, then combined after taking the modulus squared.

The third coronagraph model corresponds to the measurement error and time delay limited contrast $C_2$ of Guyon\cite{2005ApJ...629..592G}, combined with the contrast from uncorrected amplitude $C_1$ due to atmospheric scintillation.  In Appendix \ref{app:guyon_contrast} we present a derivation of $C_2$ utilizing our present notation and generalizing it for arbitrary parameters.  This method uses the residual PSD, as in Equation (\ref{eqn:var_naive}), and so requires a convolution.  Hence this tests the convolution solution for the discrepancy between the true variance and the simple PSD estimate highlighted above.

The fourth model consisted of directly measuring the variance of each modal coefficient.  The coefficient of each mode was measured by inner product with the mean-subtracted phase and amplitude on the aperture for each randomly generated screen.  In this case, the aperture is already taken into account so Equation (\ref{eqn:contrast_h2}) describes the results.

The final comparison coronagraph was an Apodized Pupil Lyot Complex Mask Coronagraph (APLCMC\cite{2014ApJ...780..171G}).  We included this as an example of a coronagraph which can be implemented in the real world.  In this case, the focal plane mask (FPM) was 1.29 $\lambda/D$ in diameter at 800 nm, with a complex transmission of -0.458.  The optimum pupil apodization gave a net intensity throughput of 0.587.

We generated a series of phase screens from the von K\'arm\'an PSD using Fourier-space convolution of the PSD with Gaussian white noise fields. Scintillation was included by multiplying the PSD by the $X$ and $Y$ functions, Equation (\ref{eqn:XandY}), resulting in phase and amplitude screens.  AO correction of the phase was approximated as in Equation (\ref{eqn:simple_est}).  We used a Fried parameter of $r_0 = 0.2$ m, an outer scale $L_0 = 25$ m, and windspeed of $\mathpzc{V} = $ 10 m/s.   The total time delay was set to 2.5 frames at 2000Hz (1.25 ms).  We set telescope diameter to $D = 6.5$ m and actuator spacing to $d=13.5$ cm (48 across), and assumed a wavelength of $\lambda=800$ nm.  The phase screens were 2048 pixels across, and we used a circular unobstructed aperture 128 pixels across.  We assumed an arbitrarily bright star so WFS measurement noise was ignored.  Strehl ratio of a non-coronagraphic image with these parameters was $\sim$94\%.  

10,000 realizations of the AO corrected phase and uncorrected amplitude were generated.  For each realization we applied the five comparison coronagraph models.  The results for these coronagraphs are shown in Figure \ref{fig:coronVerification}, where we see that, in general, these various coronagraph models match the APLCMC.  We note that the ``perfect Coronagraph'' and direct propagation using Equation (\ref{eqn:perrin_psf}) are nearly identical to each other.  The residual-PSD+convolution analysis also matches well. So does the direct calculation of modal variance.  

\subsection{Long-Exposure Contrast and the PSD}

We can summarize the results of this section with the following points:
\begin{itemize}
\item The post-coronagraph contrast is not simply the residual PSD, due to the effect of the aperture.  This is easily addressed by an unnormalized convolution with the PSF. 
\item If the variance is calculated with the aperture included, then the convolution is not needed.
\item Ideal coronagraph models are valid in the regime of ground-based, well-corrected, AO-fed coronagraphy, and are useful proxies for real-world coronagraphs.
\end{itemize}

Our result that the post-coronagraph long exposure image is a convolution of the PSF with the residual PSD is nearly identical to that developed by Herscovici-Schiller et al. \cite{2017MNRAS.467L.105H} It is fully equivalent if one uses a position dependent PSF and includes internal aberrations, both details we neglect since we are analyzing an ideal system.  The difference between this case, and the case when the aperture is included, is important when the modal variances are individually calculated as we do in the following sections.

\section{Temporal Power Spectra}
\label{sec:temporal_psds}

Next we describe a process for calculating the temporal PSDs of the Fourier basis in wind-driven von \karman turbulence.

\subsection{Temporal PSD of Fourier Modes in Wind-Driven Turbulence}
\label{sec:temp_psd_wind}

Re-stating Equation (\ref{eqn:psd_spatial}), the spatial PSD of a Fourier mode on an unobstructed circular aperture is given by\footnote{As noted above, we are suppressing $p$}:
\begin{equation}
\mathcal{P}_{mn}(\vec{k}) = \mathcal{P}_o(\vec{k}) \left|  {Q}_{mn} (\vec{k})   \right|^2
\label{eqn:psd_fourier_spatial}
\end{equation}
We next invoke the Taylor or frozen-flow hypothesis, which for our purposes states that we can treat the time-domain behavior of turbulence as if fixed turbulent phase screens are blowing across the telescope in discrete layers.  With this assumption we can calculate the temporal PSD of a turbulent mode with \cite{1993JOSAA..10..957R, 1995JOSAA..12.1559C}
\begin{equation}
\mathcal{T}_{mn}(f;\mathpzc{V}) = \frac{2}{\mathpzc{V}} \int_{-\infty}^{\infty} \mathcal{P}_{mn}(\frac{f}{\mathpzc{V}}, k_v) \mathrm{d} k_v
\label{eqn:psd_temporal}
\end{equation}
where $f$ is temporal frequency, $\mathpzc{V}$ is the wind-speed, and $\mathcal{T}_{mn}$ denotes the temporal PSD of the mode specified by $mn$. With no loss of generality we have chosen coordinates such that the wind vector $\vec{\mathpzc{V}} = \mathpzc{V}\hat{\mathpzc{u}}$.  We have included a factor of 2 since we will consider only $ f > 0$.  

Now to account for an arbitrary wind direction, let $\Theta$ be the direction of the wind with respect to the $\hat{\mathpzc{u}}$ axis.  We then derotate the basis functions by this angle to align with the axes of the integral in Equation (\ref{eqn:psd_temporal}) by defining
\begin{align}
\begin{split}
m' = & m\cos(\Theta) - n\sin(\Theta)\\
n' = & m\sin(\Theta) + n\cos(\Theta).
\end{split}
\end{align}
The temporal PSD for arbitrary wind direction is then simply
\begin{equation}
\mathcal{T}_{mn}(f;\mathpzc{V}, \Theta) = \frac{4}{\mathpzc{V}} \int_{-\infty}^{\infty} \mathcal{P}_{m'n'}(\frac{f}{\mathpzc{V}}, k_v) \mathrm{d} k_v.
\label{eqn:psd_temporal_arb}
\end{equation}

Finally, the temporal PSD of a Fourier mode for a turbulence profile is the $C_\mathpzc{n}^2$ weighted sum of the single layer profiles
\begin{equation}
\mathcal{T}_{mn}(f)  =  \sum_i C_{\mathpzc{n}}^2(z_i) \mathcal{T}_{mn}(f;\mathpzc{V}_i, \Theta_i). 
\label{eqn:psd_temporal_cn2}
\end{equation}

\subsection{Numerical Calculation}
\label{sec:temp_psd_wind_calc}

The improper integral in Equation (\ref{eqn:psd_temporal_arb}) does not in general have a simple closed-form solution, so we must calculate it numerically.  In this work we used the GSL { \verb gsl_integration_qagi }\footnote{\url{https://www.gnu.org/software/gsl/manual/html_node/QAGI-adaptive-integration-on-infinite-intervals.html}} routine.  We found that a relative tolerance of $10^{-4}$ and an absolute tolerance of $10^{-10}$ in {\verb double } precision gave good results. Though these choices noticeably extend calculation time, less demanding tolerances can produce numerically unsatisfying results.  For example, modes with similar spatial frequencies should have similar total variance (integral of the PSD), but setting numerical tolerance too low can introduce occasional 50\% jumps.

\begin{table}
\centering
\footnotesize
\caption{Las Campanas Observatory turbulence profile for median conditions assumed in this work.  Based on the GMT site survey \cite{prieto_2010SPIE,2011arXiv1101.2340T}. 
\label{tab:profile_lco}}
\begin{tabular}{lcccc}
Layer & Layer       &  Layer         & Wind           &  Wind \\
 No.  & Height$^1$  & Strength$^2$   & Dir$^3$            &  Speed \\
      & $z$ [m] & $C_n^2$  & $\Theta$ [deg] &  $\mathpzc{V}$ [m/s] \\ 
\hline
\hline
1 & 250   &  0.42    & 60   &  10 \\
2 & 500   &  0.03   & 60   &  10 \\ 
3 & 1000  &  0.06   & 75   &  20 \\
4 & 2000  &  0.16    & 75   &  20 \\
5 & 4000  &  0.11    & 100  &  25 \\
6 & 8000  &  0.10    & 110  &  30 \\
7 & 16000 &  0.12    & 100  &  25 \\
\hline
\multicolumn{5}{l}{Notes:}\\
\multicolumn{5}{l}{$^1$Height above the observatory.}\\
\multicolumn{5}{l}{$^2$Normalized.}\\
\multicolumn{5}{l}{$^3$Angle east of North.}\\
\end{tabular}
\end{table}

% \begin{table}
% \centering
% \footnotesize
% \caption{Mauna Kea Observatory turbulence profile for median conditions assumed in this work.  Based on data provided by M. Chun (private communication).
% \label{tab:profile_mko}}
% \begin{tabular}{lcccc}
% Layer & Layer       &  Layer         & Wind           &  Wind \\
%  No.  & Height$^1$  & Strength$^2$   & Dir$^3$            &  Speed \\
%       & $z$ [m] & $C_n^2$  & $\Theta$ [deg] &  $\mathpzc{V}$ [m/s] \\ 
% \hline
% \hline
% 1 &  52  &  0.69    & 90   &  6.5 \\
% 2 &  807  &  0.07   & 90   &  6.5 \\ 
% 3 & 4349  &  0.10   & 260   &  12.8 \\
% 4 & 8000  &  0.06    & 260   &  22 \\
% 5 & 12000  &  0.05    & 260  &  9.5 \\
% 6 & 19501  &  0.03    & 260  &  5.6 \\
% \hline
% \multicolumn{5}{l}{Notes:}\\
% \multicolumn{5}{l}{$^1$Height above the observatory.}\\
% \multicolumn{5}{l}{$^2$Normalized.}\\
% \multicolumn{5}{l}{$^3$Angle east of North.}\\
% \end{tabular}
% \end{table}
We also find that temporal frequency sampling is important, settling on $\Delta f = 0.1$ Hz as the largest sampling that should be used in general.  Larger samplings tend to also produce jumps, caused by moving on and off sharp peaks associated with individual wind-layers.  If such a peak falls between two frequencies and the sampling is too coarse, the power in the peak will be missed.

\begin{figure}[t]
\footnotesize
\centering
\includegraphics[width=6.5in]{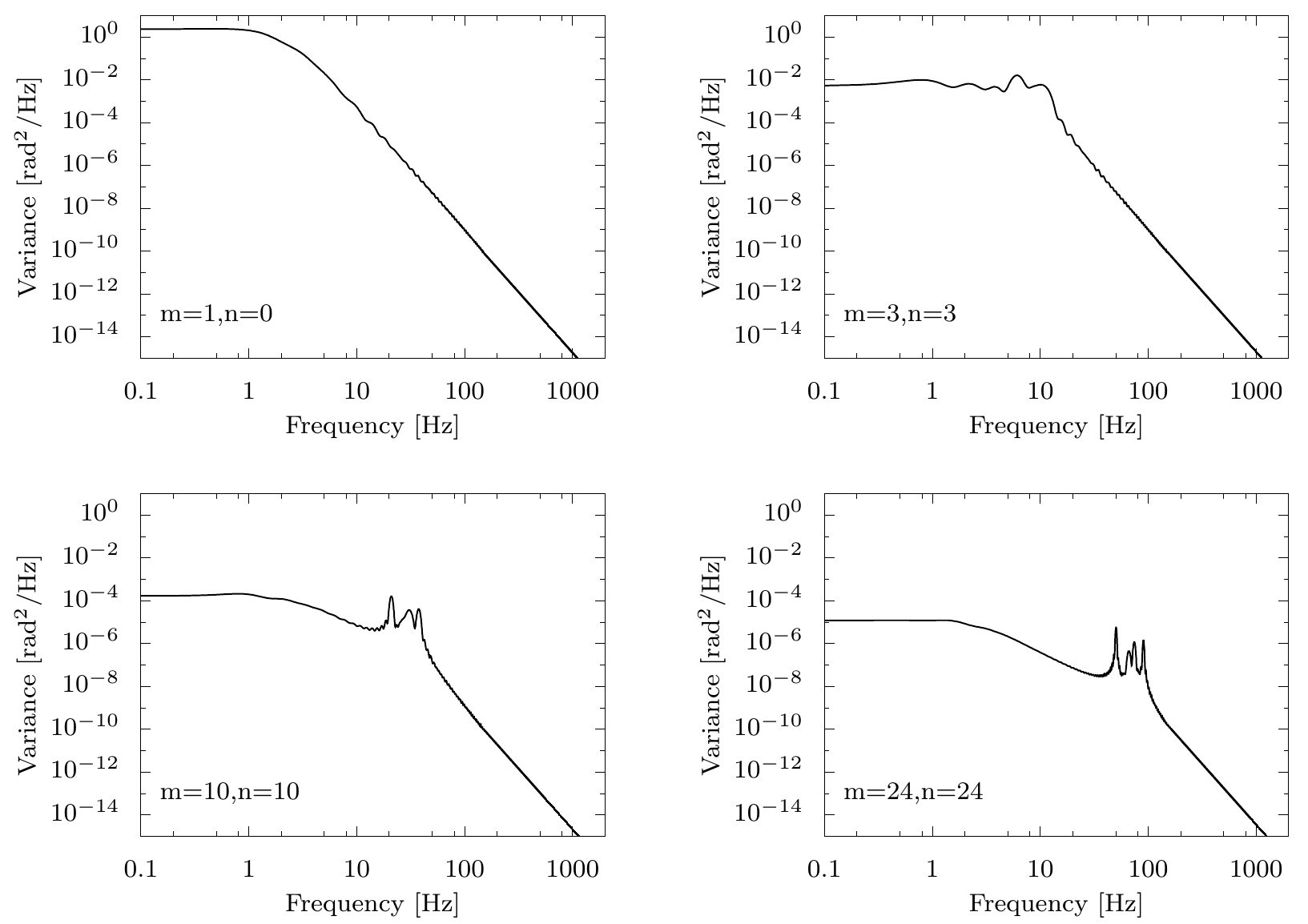}
\caption{Calculated temporal PSDs in a 7-layer turbulence model at 4 different modes, corresponding to discrete spatial frequencies. \label{fig:psdsimcomp}}
\end{figure}

We illustrate calculations of multi-layer turbulence using a seven layer turbulence model based on the Giant Magellan Telescope (GMT) site survey at Las Campanas Observatory (LCO) \cite{prieto_2010SPIE, 2011arXiv1101.2340T}.  The parameters of this profile are given in Table \ref{tab:profile_lco}.  The calculated temporal PSDs for several Fourier modes are shown in Figure \ref{fig:psdsimcomp}.   Due to the identical wind velocities of several layers, we only expect 4 distinct peaks.  The number of peaks seen at a given spatial frequency depends on $f_{pk,i} = \vec{\mathpzc{V}}_i \: \cdot\: \vec{k}_{m,n}$.  If the frequency of two peaks are not well separated they appear blended at the resolution of the calculations and plots.

\subsection{Temporal PSD of Measurement Noise}
\label{sec:psd_phn}
We also need the PSD of the WFS measurement noise. The variance at a spatial frequency due to photon noise in the wavefront sensor is\cite{2005ApJ...629..592G} 
\begin{equation}
\sigma_{ph,mn}^2 = \frac{ \beta_p^2(\vec{k}_{mn}) }{N_{ph}} \mbox{ [rad}^2\mbox{ rms]} 
\label{eqn:sigma_ph_f}
\end{equation}
where the factor $\beta_p(\vec{k}_{mn})$ quantifies the sensitivity of the WFS to photon noise.   $N_{ph}$ is the total number of photons available for wavefront sensing:
\begin{equation}
N_{ph} = F_{\gamma} \tauwfs
\end{equation}
where $F_{\gamma}$ is the photon rate (photons sec$^{-1}$) sensed by the WFS, and $\tauwfs$ is the integration time of the WFS.  $F_{\gamma}$ depends on star brightness through
\begin{equation}
F_{\gamma} = F_{\gamma,0} 10^{-0.4 \times mag}
\end{equation}
Where $mag$ is the Vega magnitude of the star in the WFS bandpass, and $F_{\gamma,0}$ is the photon rate of a 0 $mag$ star in the WFS bandpass including total system throughput. 
Here we expand this to include background and readnoise. The signal-to-noise ratio is
\begin{equation}
S/N = \frac{ F_{\gamma} \tauwfs }{\sqrt{ F_{\gamma} \tauwfs + n_{px} F_{bg}\tauwfs + n_{px}\sigma^2_{ron} } }
\end{equation}
where $n_{px}$ is the number of pixels involved in measuring $N_{ph}$, each pixel having readout noise with variance $\sigma^2_{ron}$, and $F_{bg}$ is the photon rate per pixel from the background.  Now the WFE due to photon noise at the WFS wavelength can be written as
\begin{equation}
\sigma_{ph,mn}^2 = \frac{ \beta_p^2(\vec{k}_{mn}) }{(S/N)^2} \mbox{ [rad}^2\mbox{ rms]}
\label{eqn:sigma_ph_f_sn}
\end{equation}
This variance is spread over the sampling bandwidth of the WFS, $1/2\tauwfs$.  We can determine the temporal PSD of measurement noise from
\begin{equation}
\int_0^{1/2\tauwfs} \mathcal{T}_{ph,mn}(f) df = \sigma_{ph,mn}^2
\end{equation}
assuming white noise (a flat PSD) we have
\begin{equation}
\mathcal{T}_{ph,mn}(f) = \sigma_{ph,mn}^2 \tauwfs
\end{equation}
Note that in the absence of detector noise, $\mathcal{T}_{ph,mn}(f)$ is independent of $\tauwfs$.

\section{Closed Loop Control}
\label{sec:controls}

Now that we have established the temporal PSDs of the Fourier modes and of WFS measurement noise, we can use them to predict the residual wavefront variance in a closed-loop AO system. Here we follow the standard model used throughout the AO literature, and so only briefly introduce the concepts and our notation.  Note that though here we apply the following to the frozen-flow turbulence PSDs we just derived, the techniques are general and could be applied to PSDs in any basis set and to PSDs including additional error sources such as telescope vibrations and non-frozen-flow.

\subsection{Transfer Functions}
\label{sec:xfer_funcs}
First we need the error and noise transfer functions of the AO control system.  Here we closely follow the development of Poyneer et al.\cite{2016ApOpt..55..323P}, and see also the treatment by Madec\cite{1999aoa..book.....R_ch6}. Our goal is to calculate the error transfer function (ETF) of the control system, which is defined as
\begin{equation}
\left|\mbox{ETF}(f)\right|^2 = \frac{ \mathcal{T}_{mn,\mbox{out}}(f) }{ \mathcal{T}_{mn,\mbox{in}}(f) }.
\end{equation}
That is, the ETF describes the AO system as a temporal filter which acts on the input PSD.  The noise transfer function (NTF) is similarly defined, except that it acts on the measurement noise PSD.  The ETF and NTF are constructed using the frequency responses, or transfer functions, of the various components of the system, which we will denote as $H$.

The WFS measurement is an average of the turbulence over the integration time.  Likewise, the deformable mirror (DM) is assumed to move in discrete steps, holding its position between updates.  The frequency response of such components is called a sample and hold, which has the functional form
\begin{equation}
H_{wfs}(s) = H_{dm}(s) = \frac{1 - e^{-sT}}{sT}
\label{eqn:H_wfs}
\end{equation}
where $T = 1/f_s$, $f_s$ being the loop sampling frequency, and $s = i 2\pi f$.  There will be a delay $\tau$ due to sensor readout, data transfer, calculations, etc.\footnote{It is common to discuss the ``total delay'' including 1/2 the WFS sample and hold (for the mid-point of the integration), and 1/2 the DM  sample and hold. That is the total delay is $T+\tau$\cite{1999aoa..book.....R_ch6}.}, which has a transfer function
\begin{equation}
H_{\tau}(s) = e^{-s\tau}
\label{eqn:H_tau}
\end{equation}

The system forms an estimate of the coefficient of a mode, $\widetilde{h}$. In closed-loop the WFS measures the change in the coefficient of a given mode, $\Delta h$.  This is combined with the current estimate of the total coefficient to form a new estimate of $\widetilde{h}$, which is then applied to the DM.  At time-step $t_i$ the coefficient is specified in terms of the  current and previous $\Delta h$ and $\widetilde{h}$, by a linear combination 
\begin{equation}
\widetilde{h} (t_i) = \sum_{j=1}^J a_j \widetilde{h}(t_{i-j}) + g\sum_{l=0}^L b_l \Delta h (t_{i-l})  
\label{eqn:linfilt}
\end{equation}
where $g$ is the loop gain. Below we consider in depth how to choose the parameters of this control filter.  The frequency response of this general linear filter is \cite{oppenheim2011discrete}
\begin{equation}
H_{con}(z;g) = \frac{g \sum_{l=0}^L b_l z^{-l}}{1+\sum_{j=1}^J a_l z^{-j} }
\label{eqn:H_con}
\end{equation}
and we can map from the $z$-domain to the $s$-domain by the substitution $z \rightarrow e^{sT}$.

With the above component frequency responses, we can write the open-loop ETF as
\begin{equation}
\mbox{ETF}_{ol}(s;g) = H_{wfs}(s) H_{con}(s;g) H_{\tau}(s) H_{dm}(s)
\label{eqn:etf_ol}
\end{equation}
From this it follows that the closed-loop ETF is
\begin{equation}
\mbox{ETF}_{cl}(s;g) = \frac{1}{1 + \mbox{ETF}_{ol}(s;g)}
\end{equation}
The closed-loop noise transfer function (NTF) is
\begin{equation}
\mbox{NTF}_{cl}(s;g) = H_{dm}(s) H_{\tau}(s)  H_{con}(s;g)\:\mbox{ETF}_{cl}(s ; g)
\end{equation}

The residual PSD in closed-loop control is then the product of the modulus-squared ETF and the modal PSD, and the modulus-squared NTF and WFS noise PSD:
\begin{equation}
\mathcal{T}_{cl,mn}(f;g) = \mathcal{T}_{mn}(f) \left| \mbox{ETF}_{cl}(s;g) \right| ^2 + \mathcal{T}_{ph,mn}(f) \left| \mbox{NTF}_{cl}(s;g) \right| ^2. 
\end{equation}
The residual variance in this mode is just the integral of the residual PSD:
\begin{equation}
\left|\sigma_{mn}(g) \right|^2 = \int_0^{f_s}   \mathcal{T}_{cl,mn}(f;g) df.
\label{eqn:closed_loop_variance}
\end{equation}
Note that, since we started with Equation (\ref{eqn:psd_fourier_spatial}), this estimate of the variance explicitly includes the aperture and so directly describes the post-coronagraph intensity.

To employ these calculations, we next need to determine the parameters of Equation (\ref{eqn:linfilt}) and then find the optimum gain.

\subsection{Control Laws}
\label{sec:control_laws}

Our next step in analyzing a closed-loop AO system is to choose the control law used to feed WFS measurements back to the DM.  Here we analyze two: the simple integrator and the Linear Predictor.

\subsubsection{Simple Integrator}
At time step $t_i$, the WFS measures the residual wavefront, $\Delta h_i$.  We combine this with the last estimate, $\widetilde{h}_{i-1}$, which is the result of the previous cycle.  The simplest estimate at the current step is the simple integrator (SI)
\begin{equation}
\widetilde{h}_i = \widetilde{h}_{i-1} + g \Delta h_i.
\end{equation}
This corresponds to a choice of $J=1;a_1 = 1$ and $L=0;b_0 = 1$ in Equation (\ref{eqn:linfilt}).  It is also common to choose $a_1 < 1$ (slightly), a so-called leak term, making this the ``leaky integrator'' control law.  Here we will confine our analysis to the SI control law.

\subsubsection{Linear Predictor} 
In general the estimate formed by the SI controller will be slightly incorrect when actually applied due to the finite loop delay.  Here we consider a conceptually straightforward extension of the integrator to include a prediction.  We note that more sophisticated methods exist, which we discuss after presenting our results.

One way to improve the accuracy of the estimate formed by the controller is with a filter of the form 
\begin{equation}
\widetilde{h}_i = \sum_{j=0}^{N-1} c_j \left(\widetilde{h}_{i-j-1} + \Delta h_{i-j}\right).
\label{eqn:lp_filter}
\end{equation}
We can determine a set of $c_j$ which are optimal in the least-squares sense from the temporal autocorrelation of the modal amplitude, $\mathcal{R}_{mn}(\mathpzc{t})$, where $\mathpzc{t}$ is the lag, by solving the Normal or {\it Yule-Walker} equations \cite{vaidyanathan2008theory, oppenheim2011discrete}:
\begin{equation}
\mathbf{R} \vec{c} = -\vec{r}.
\end{equation}
where
\begin{equation}
\mathbf{R} = 
\begin{bmatrix}
\mathcal{R}_{mn}(0) & \mathcal{R}_{mn}(1)                  & \cdots & \mathcal{R}_{mn}(N-1) \\
\mathcal{R}_{mn}(1)                   & \mathcal{R}_{mn}(0)           & \vdots & \mathcal{R}_{mn}(N-2) \\
\vdots                   &  \hdots                & \ddots & \vdots \\
\mathcal{R}_{mn}(N-1)                 & \mathcal{R}_{mn}(N-2)                  & \cdots & \mathcal{R}_{mn}(0)
\end{bmatrix}
\end{equation}
and
\begin{equation}
\vec{r} = 
\begin{bmatrix}
\mathcal{R}_{mn}(1)\\
\mathcal{R}_{mn}(2)\\
\vdots\\
\mathcal{R}_{mn}(N)
\end{bmatrix}
\end{equation}
We can find the autocorrelation of the time-series of a mode from its PSD using the Wiener-Khinchin theorem \cite{honerkamp1993stochastic}, which states that the autocorrelation is the Fourier transform of the PSD:
\begin{equation}
\mathcal{R}_{mn}(\mathpzc{t}) = \mathcal{F} \left\{ \mathcal{T}_{mn} (f) \right\}
\end{equation}
In Appendix \ref{app:autocorr} we give a recipe for calculating the autocorrelation from a numerical PSD.

The solution vector $\vec{c} = \left[c_0, c_1, \cdots, c_{N-1} \right]$ contains the optimal {\it Linear Predictor} (LP) coefficients, which minimize the least-squares error of the output of Equation (\ref{eqn:lp_filter}).  The matrix $\mathbf{R}$ is a symmetric Toeplitz matrix, and so the Normal equations can be solved very efficiently using Levinson's recursion \cite{1992nrca.book.....P}(chapter 2). A nice result of using Levinson's recursion to solve the Yule-Walker equations is that the resultant filter is guaranteed to be stable \cite{vaidyanathan2008theory}. 

To apply the LP in closed-loop, we use $\vec{c}$ as the coefficients in Equation (\ref{eqn:linfilt}).  That is, $L = N-1$, $b_l = c_{l}$ and $J = N$, $a_j = c_{j-1}$.  We can then to determine the optimal loop gain $g$.

\subsection{Gain Optimization}
\label{sec:gain_opt}
\begin{figure}
\centering
%\scalebox{0.8}{\input{figures/nyquist}}
\includegraphics[width=4in]{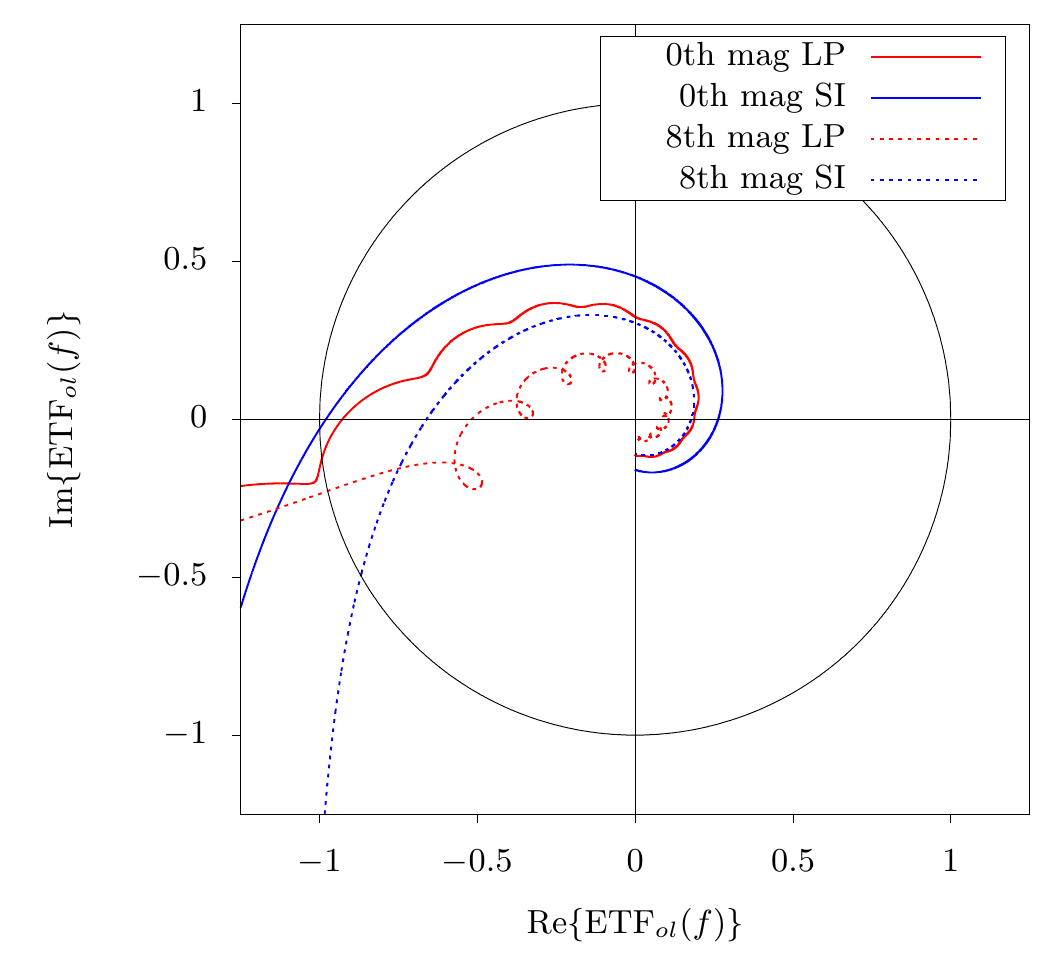}
\caption{Nyquist plot demonstrating the stability of our two control laws.  The open-loop ETFs are plotted in the imaginary plane for both the SI and LP control laws.  Note that frequency increases as the lines move out from the center.  The controllers are stable so long as they do not encircle (0,-1). The value of gain $g$ which causes them to intersect (0,-1), and hence become unstable, is here called the maximum stable gain $g_{max}$.  The black circle has a radius of 1.  We show the open-loop ETFs for each controller, with optimum gains for both a 0th and an 8th magnitude star.
\label{fig:nyquist}}
\end{figure}

All that remains is to describe how to choose the optimum value of the gain.  The first step is to identify the maximum stable gain, $g_{max}$, which is the value of $g$ at which the system becomes unstable\footnote{This is sometimes called the ``ultimate gain''.}.  Following Dessenne et al.\cite{1998ApOpt..37.4623D} we evaluate stability of the control law using a Nyquist plot of the open-loop ETF, which plots the real and imaginary components of Equation (\ref{eqn:etf_ol}).  In this plane, as long as the ETF does not encircle the point (0,-1), then the system is stable -- a result known as Nyquist's stability criterion.    We then can find the value of $g_{max}$ by finding the value of $g$ which causes the curve to intersect (0,-1).  This is illustrated in Figure \ref{fig:nyquist}.

From here we follow the standard recipe \cite{1994A&A...291..337G, 2005JOSAA..22.1515P} for using the input (or ``open-loop'') PSD to optimize the closed-loop gain.  The per-mode variance-minimizing optimum gain is given by solving:
\begin{equation}
g_{opt} = \argmin_{0 \le g < g_{max}} \left|\sigma_{mn}(g) \right|^2. 
\end{equation}
This is a straightforward optimization problem, which we solve using Brent's method as implemented in the boost c++ library.

Integrator control with optimal gains is implemented in the SPHERE SAXO system, which uses a Karhunen-Lo\'{e}ve basis with optimal modal gains \cite{2014SPIE.9148E..0OP}.  The GPI instrument likewise uses optimal gains but with Fourier modes \cite{2016ApOpt..55..323P}.

\section{Closed Loop Contrast Calculations}
\label{sec:calcs}

Given the temporal PSD, the ETF and NTF, and the optimal gain, equation (\ref{eqn:closed_loop_variance}) describes the residual variance at a given spatial frequency in a closed-loop AO system.  We showed how this is related directly to the PSF contrast behind a coronagraph in Section \ref{sec:psf_contrast}.  So, calculating the closed-loop long-exposure post-coronagraph contrast consists of the following steps:
\begin{enumerate}
\item Calculate the temporal PSDs at each $(m,n)$ according to Sections \ref{sec:temp_psd_wind} and \ref{sec:temp_psd_wind_calc}.  Due to the Hermitian symmetry of the Fourier basis this only needs to be done in half the plane.
\item Calculate the temporal PSD of measurement noise according to Section \ref{sec:psd_phn}.
\item Determine the optimum controller coefficients per Section \ref{sec:control_laws}, and the associated optimum gains per Section \ref{sec:gain_opt}.
\item Determine the ETF and NTF per Section \ref{sec:xfer_funcs}
\item Populate the residual PSD as follows:
\begin{equation}
\sigma_{mn}^2 =  
\begin{cases}
\left|\sigma_{mn}(g_{opt}) \right|^2, & \mbox{if } g_{opt} > 0.\\
\mathcal{P}_{mn}(\vec{k})/D^2, & \mbox{if } g_{opt} = 0.
\end{cases}
\end{equation}
\item Calculate the intensity:
\begin{enumerate}
\item For the corrected modes, the variance is properly sampled and directly describes the intensity according to Equation (\ref{eqn:contrast_h2}).
\item For the uncorrected modes, we must use Equation (\ref{eqn:contrast_var_mn}) and then (\ref{eqn:contrast_Phi}).  Note that we include the corrected-mode variances in the convolution.
\end{enumerate}
\item Calculate the long-exposure Strehl ratio as the sum of variance over all spatial frequencies to find $\sigma^2_{tot}$ and use the extended Marechal approximation
\begin{equation}
\left< S \right> = \mbox{e}^{-\sigma^2_{tot}}.
\end{equation}
\item Calculate the contrast according to Equation (\ref{eqn:psf_contrast}).
\end{enumerate}

\begin{figure}[h!]
\centering
\footnotesize
\includegraphics[width=6.5in]{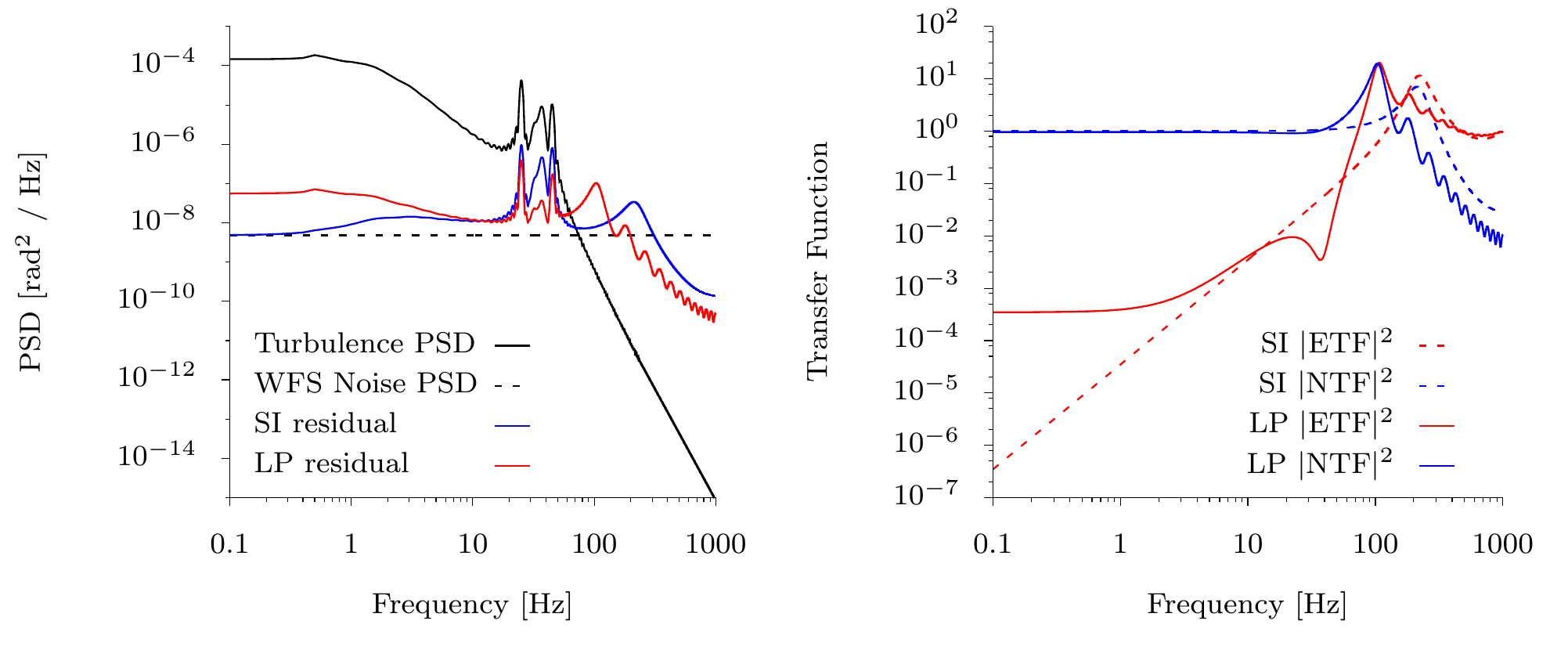}
\caption{ Left: the input turbulent PSD for Fourier mode $(m=12,n=12)$ (black), along with the residual PSDs in closed-loop control for the SI (blue) and LP (red) controllers.  The WFS Noise PSD (dashed) corresponds to an 8th mag star on a 6.5 m telescope.
Right: The ETF (solid) and NTF (dashed) for both SI (blue) and LP (red) controllers, optimized for the input PSDs shown at left.  \label{fig:xfer_funcs}}
\end{figure}

The first four steps are illustrated in Figure \ref{fig:xfer_funcs}. The PSD of a Fourier mode is shown in the left-hand panel.  The gain-optimized ETFs and NTFs, for both SI and LP controllers, are shown in the right-hand panel.  The left-hand panel also shows the residual PSDs after applying the ETFs and NTFs to the input turbulence PSD. 

\begin{figure}[h!]
\centering
\footnotesize
\includegraphics[width=6in]{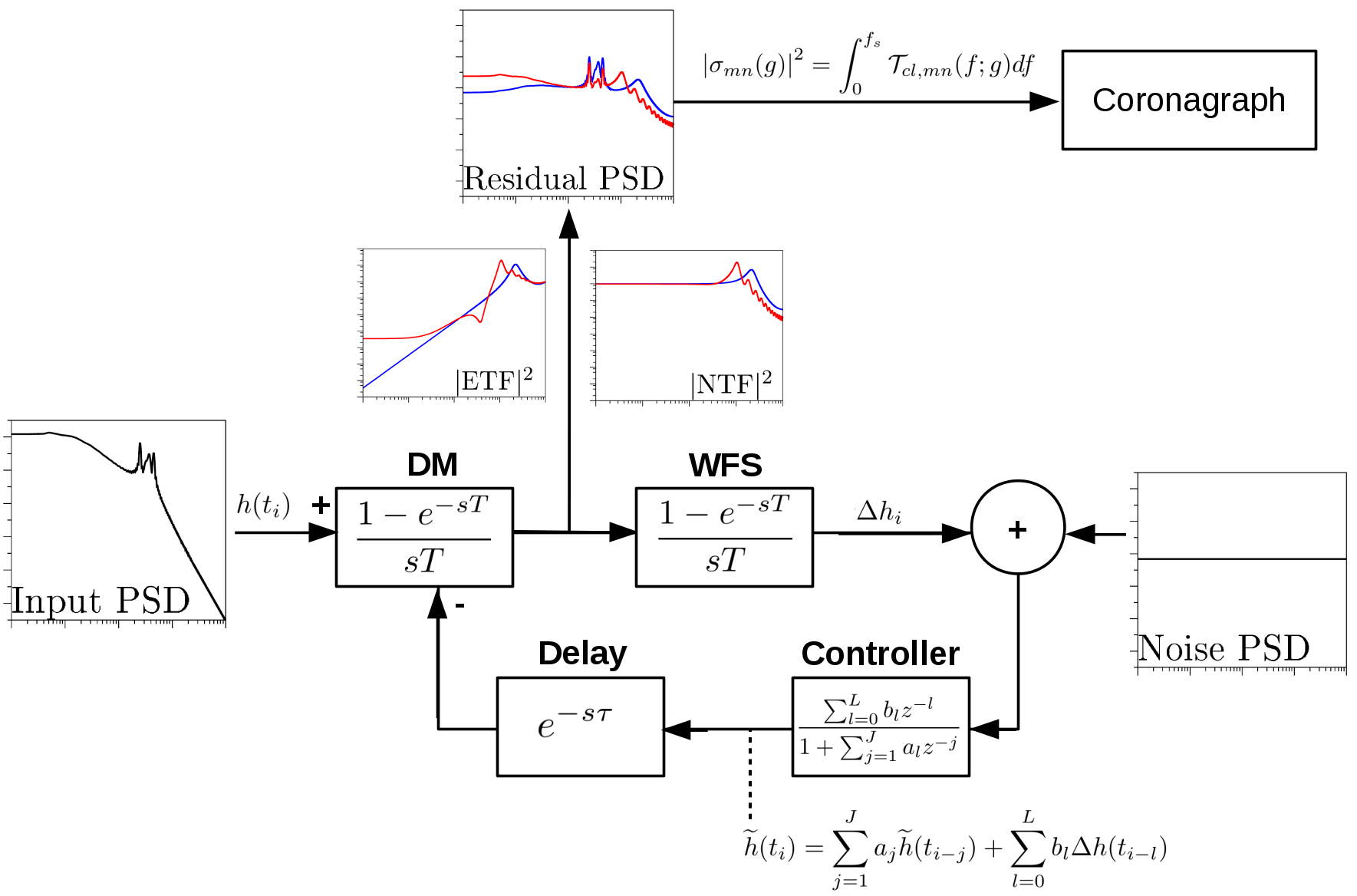}
\caption{Illustration of our framework, using one input PSD as an example.  The input wavefront, described by its temporal PSD, is corrected by the DM.  The WFS measures the residual, and noise is added as described by a temporal PSD.  A controller, represented as a linear filter, determines an estimate of the wavefront at the next time-step. After a delay, this estimate determines the new shape of the DM.  The control loop acts on the input PSD according to the error transfer function (ETF) and on the noise PSD according to the noise transfer function (NTF).  The resultant residual temporal PSD determines the residual variance of this mode, which is then input to the coronagraph. Blue curves are for the simple integrator, and red curves are for the linear predictor.
\label{fig:loopFig}}
\end{figure}

 We illustrate the complete framework in Figure \ref{fig:loopFig}. The frequency response of each component is applied to the input PSD, and to the noise PSD, according to the ETF and NTF.  The output residual PSD is then input to the coronagraph.

We applied our method to the cases of a 6.5 m circular aperture, here motivated by the in-development MagAO-X system \cite{2016SPIE.9909E..52M}.  We also consider a 25.4 m telescope, for which we used the collecting area of the GMT (seven 8.4 m segments), but ignored details of the aperture in PSD and contrast calculations.  The parameters of our calculations for these telescopes are listed in Table \ref{tab:params}.  

\begin{table}[t]
\footnotesize
\centering
\caption{Parameters. \label{tab:params}}
\begin{tabular}{lcccl}
    Symbol           & Parameter                   &  Value             &  Units               & Notes \\
\hline
\hline
 $f_s$               & Max sampling frequency      &  2000              & Hz                   & \\
 $\tau$              & Loop delay                  &  1.5               & frames               & \\
 $d$                 & Actuator spacing            &  0.135             & m                    & 48 across 6.5 m  \\
 $\beta_p$           & WFS Sensitivity             & $\sqrt{2}$         &                      & For unmodulated pyramid   \\     
 $\lambda$           & WFS Wavelength              &  800               & nm                   & Based on the MagAO WFS \cite{2016SPIE.9909E..01M}\\     
                     & WFS Filter Width            &  357               & nm                   & Based on the MagAO WFS \cite{2016SPIE.9909E..01M}\\
                     & WFS Throughput              &  0.1               & [ratio]              &             \\
 $\tau_{wfs}$        & WFS Exposure Time           & 0.0005             & s                    & 2 kHz\\
 $\sigma_{ron}$      & WFS Readout noise           &  0.1               & electrons            &            \\
                     & 0 mag flux density @ 800 nm &  $5.0\times10^7$   & photons/sec/m$^2$/nm & Vega based. \\
                     & Sky background @ 800 nm     &  20                & mags/arcsecond$^2$   & \\
  $r_0$              &Fried Param.                & 0.16               & m                    & Median 0.62'' at $0.5$ $\mu$m \cite{2011arXiv1101.2340T} \\
 $\bar{\mathpzc{V}}$ &Weighted mean wind speed    &   18.7             & m/s                  & $5/3$ moment \cite{1998aoat.book.....H} \\                     
\hline
\multicolumn{4}{l}{Magellan Clay}\\
\hline
 $D$                 &Telescope Diameter          & 6.5                & m                    & \\
 $F_\gamma,0$        &0 mag Photon Flux           & $5.9\times10^{10}$ & photons/sec          & Unobstructed aperture \\
  $n_{px}$           &WFS Pixels                  & 7280               & pixels               & 4 quadrants with D/d across\\
  $F_{bg}$           &Background flux             & 0.22               & photons/pix/sec      & Assumes 2'' FOV for WFS\\
\hline
\multicolumn{4}{l}{GMT}\\
\hline
 $D$                 &Telescope Diameter          & 25.4               & m                    & Edge to edge \\
 $F_\gamma,0$        &0 mag Photon Flux           & $6.6\times10^{11}$ &  photons/sec         & Vega, using GMT aperture \\
  $n_{px}$           &WFS Pixels                  & 111,208            & pixels               & 4 quadrants with D/d across \\
  $F_{bg}$           &Background flux             & 0.01               & photons/pix/sec      & Assumes 2'' FOV for WFS \\
\hline
\end{tabular}
\end{table}

\begin{figure}
\centering
\footnotesize
\includegraphics[width=5in]{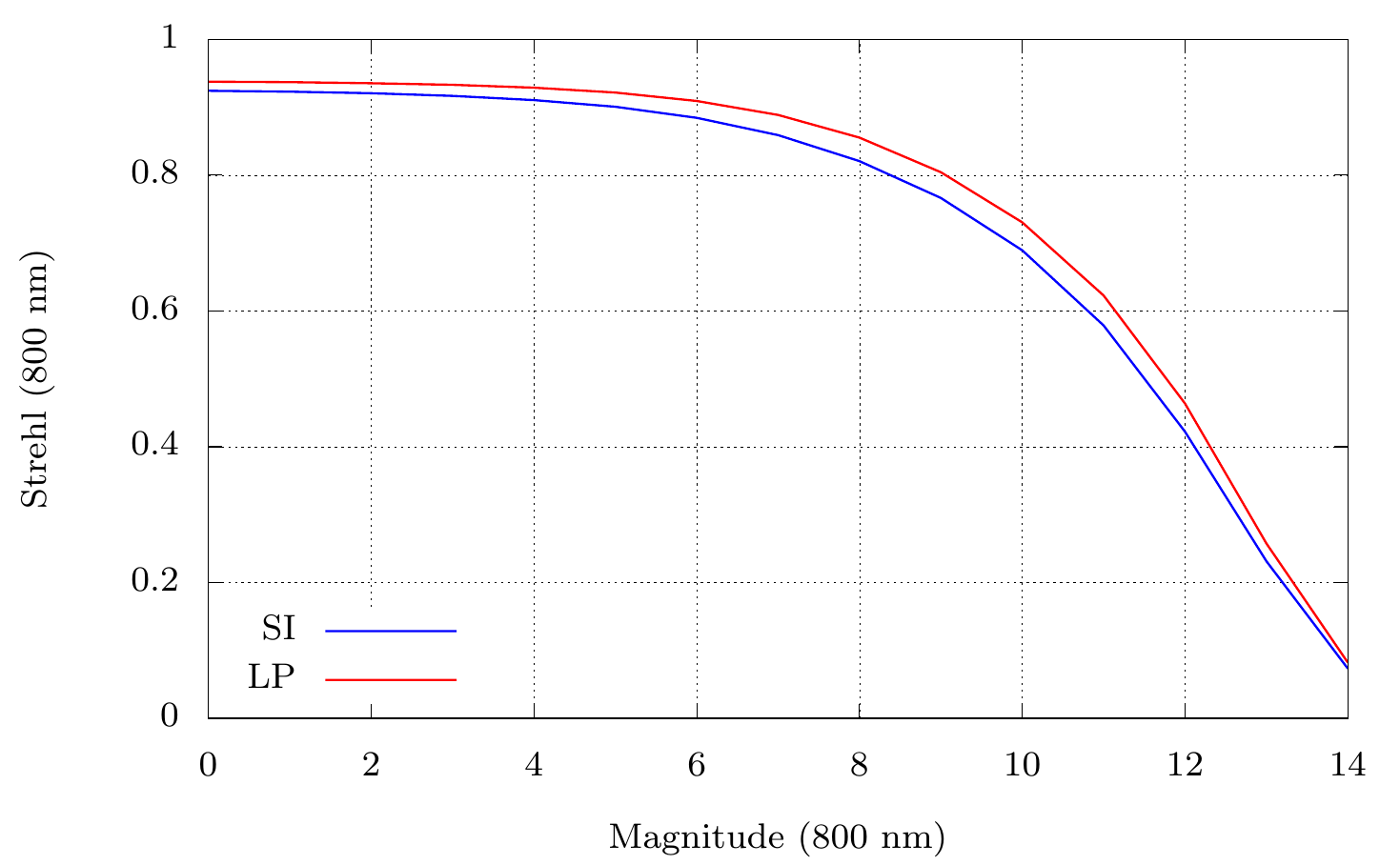}
\caption{Strehl ratio vs guide star magnitude with optimized gains for the two controllers at the WFS wavelength (800 nm). Only residual modal variance and fitting error are included in this plot. The LP controller improves Strehl at all magnitudes, with the benefit being greatest between 8th and 10th magnitude.   \label{fig:strehl} }
\end{figure}

Figure \ref{fig:strehl} shows the Strehl ratio vs. guide star magnitude predicted for the 6.5 m AO system at 800 nm.  Only residual turbulence is accounted for, including the uncorrected power (a.k.a. fitting error).

\begin{figure}
\centering
\includegraphics[width=6in]{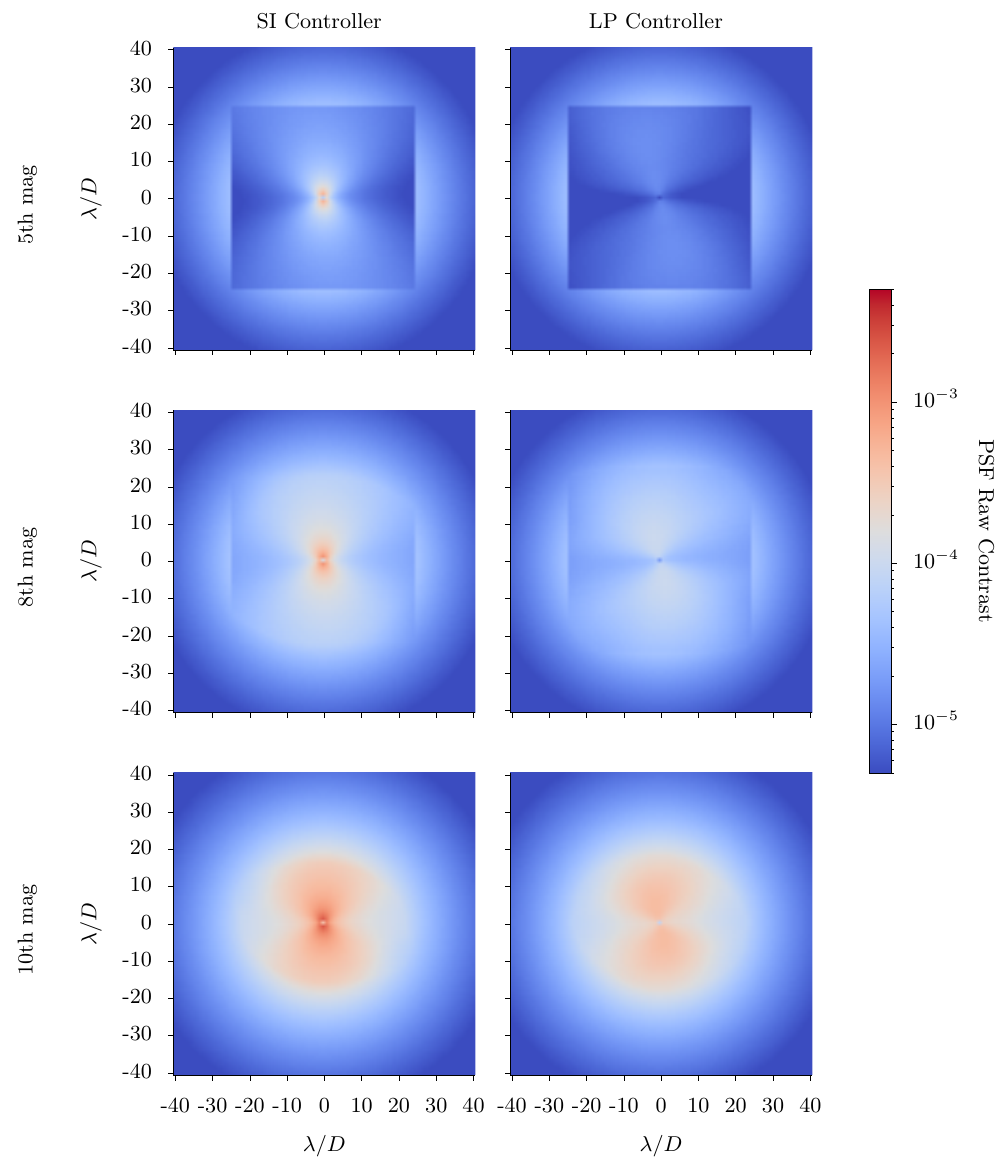}
\footnotesize
\caption{Post-coronagraph long exposure contrast maps for a 6.5 m telescope at 800 nm.  Here we compare results for the simple integrator (SI) and linear predictor (LP) controllers, for 5th, 8th, and 10th magnitude stars.  As we do not address chromaticity here, the WFS and observation wavelengths are both 800 nm.
\label{fig:cmap_6_5m_50th}}
\end{figure}

\begin{figure}
\centering
\includegraphics[width=6.5in]{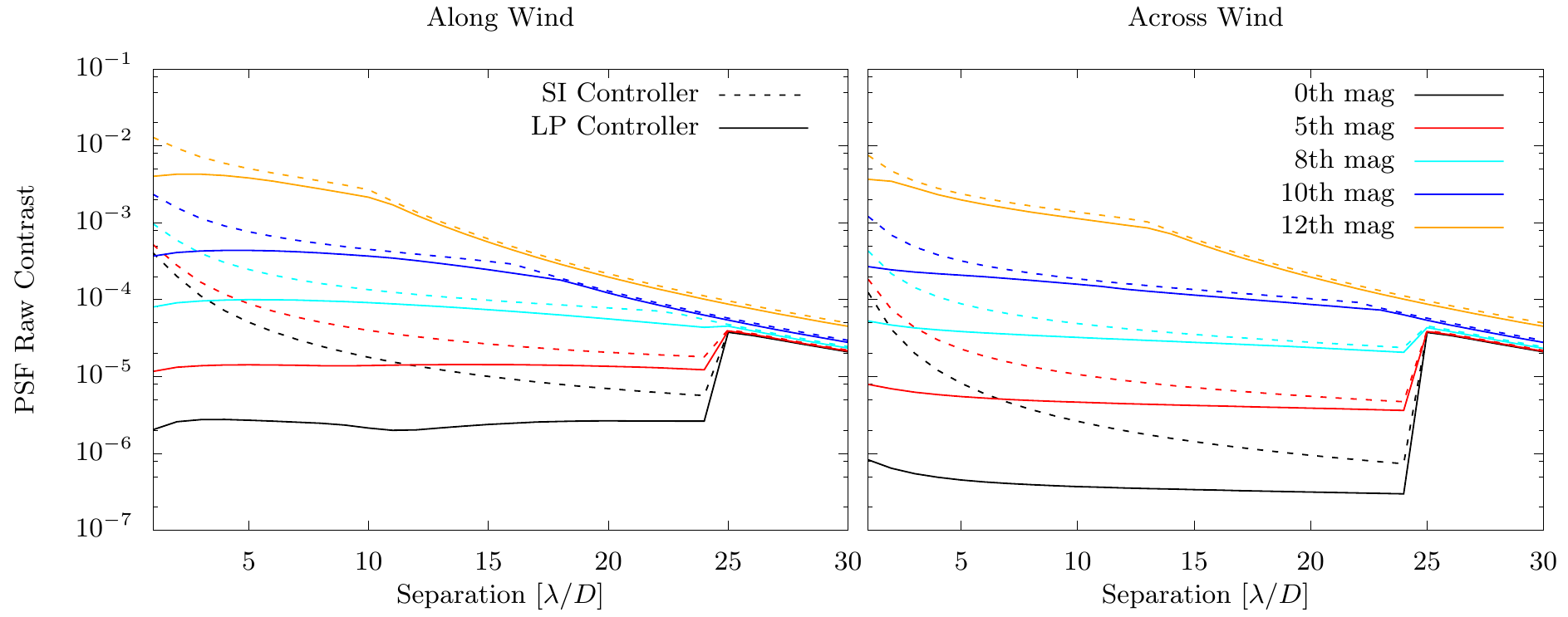}
\caption{Contrast profiles for a 6.5 m telescope observing at 800 nm, using the parameters shown in Tables \ref{tab:profile_lco} and \ref{tab:params}. As we do not address chromaticity here, the WFS and observation wavelengths are both 800 nm.  The left panel is along the main wind direction (up-and-down in Figure \ref{fig:cmap_6_5m_50th}), and the right panel is perpendicular.  The dashed lines are for the simple integrator (SI) controller, standard in current AO systems.  The solid lines are for the Linear Predictor (LP) controller, the form of predictive control we analyze in this work. Since we are conducting a mode-by-mode analysis in discrete spatial frequencies, the lowest spatial frequency we analyze is $1/D$, so the smallest separation we show is 1 $\lambda/D$. \label{fig:cprof_6_5} }
\end{figure}

In Figure \ref{fig:cmap_6_5m_50th} we show the contrast maps for the 6.5 m system observing 5th, 8th, and 10th magnitude stars, for both controllers at the WFS wavelength.  To estimate the PSF contrast at other wavelengths all of these results can be scaled by $1/\lambda^2$.  The figure shows, as expected, deeper contrasts on brighter stars.  The SI controller (left hand column) exhibits the well known ``wind butterfly'', and we see that the gain optimization process turns off modes on fainter stars, changing the shape of the dark hole.  The LP controller does not show the ``wind butterfly'', rather predictive control essentially eliminates it.  It also results in deeper contrasts. Figure \ref{fig:cprof_6_5} shows the PSF contrast profiles both along and across the wind direction.

\begin{figure}
\centering
\includegraphics[width=6in]{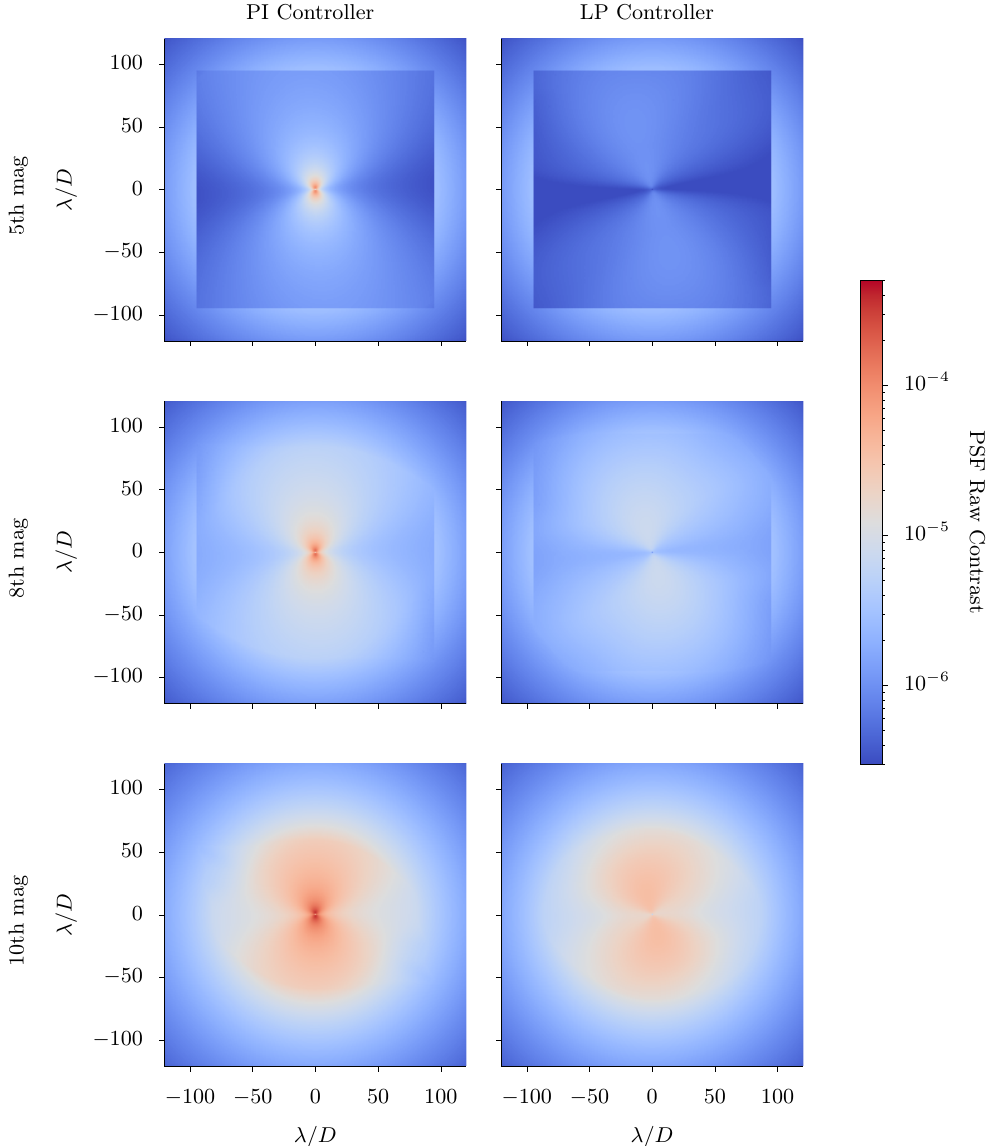}
\footnotesize
\caption{Post-coronagraph long exposure contrast maps for a 25.4 m telescope at 800 nm, with collecting area defined by the GMT aperture.  Here we compare results for the simple integrator (SI) and linear predictor (LP) controllers, for 5th, 8th, and 10th magnitude stars.  Note that the color scale is x10 lower than in Figure \ref{fig:cmap_6_5m_50th}.   As we do not address chromaticity here, the WFS and observation wavelengths are both 800 nm.
\label{fig:cmap_25_4m_50th}}
\end{figure}

\begin{figure}
\centering
\includegraphics[width=6.5in]{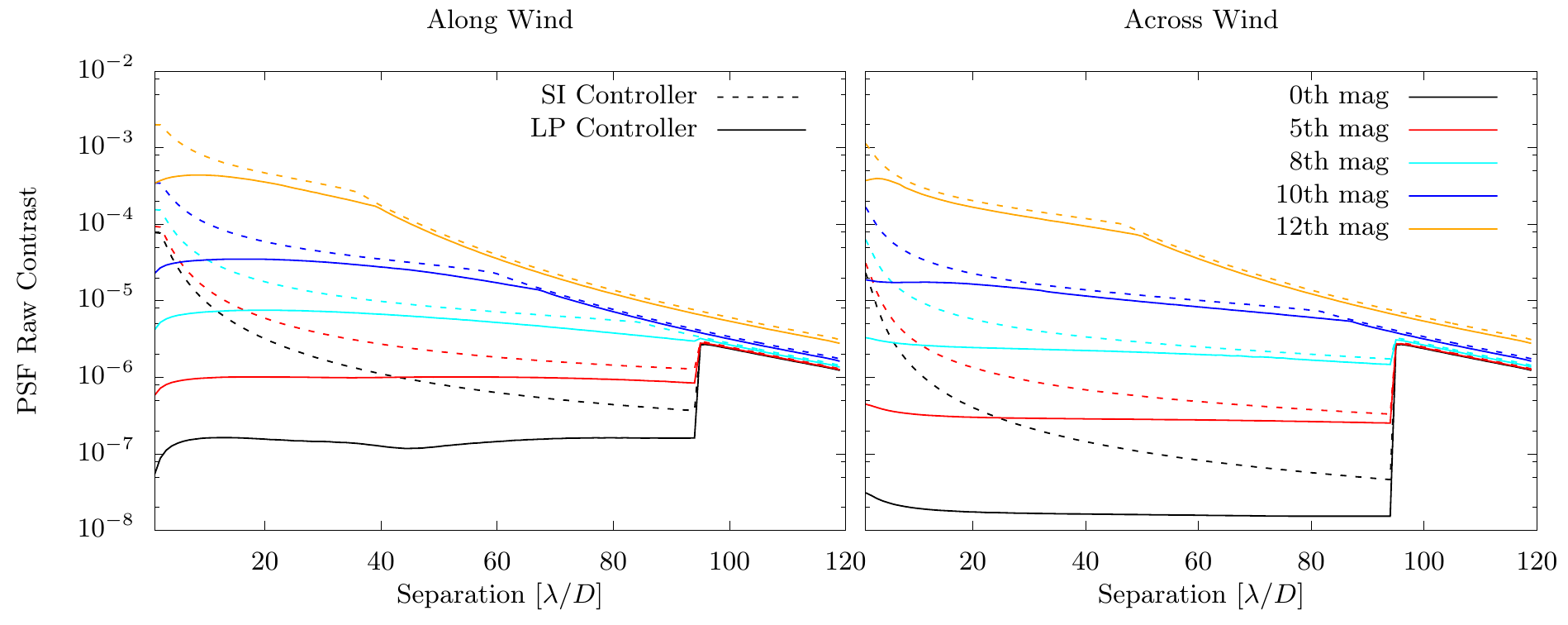}
\caption{Contrast profiles for a 25.4 m telescope with the collecting area of the GMT aperture, observing at 800 nm, using the parameters shown in Tables \ref{tab:profile_lco} and \ref{tab:params}. As we do not address chromaticity here, the WFS and observation wavelengths are both 800 nm.  The left panel is along the main wind direction (up-and-down in Figure \ref{fig:cmap_25_4m_50th}), and the right panel is perpendicular.  The dashed lines are for the simple integrator (SI) controller, standard in current AO systems.  The solid lines are for the Linear Predictor (LP) controller, the form of predictive control we analyze in this work.  Since we are conducting a mode-by-mode analysis in discrete spatial frequencies, the lowest spatial frequency we analyze is $1/D$, so the smallest separation we show is 1 $\lambda/D$. \label{fig:cprof_25_4}  }
\end{figure}

The same calculations were made for the 25.4 GMT-like telescope, shown in Figures \ref{fig:cmap_25_4m_50th} and \ref{fig:cprof_25_4}.  The results are qualitatively similar, but show a nearly $\times10$ improvement due to the larger aperture.  This can be understood as the larger diameter causing the spatial-frequency sampling interval, $1/D^2$, to impose less power per $\lambda/D$,  resulting in less residual power per mode.

For both the 6.5 m and the 25.4 m telescopes, the gain in contrast at small separations provided by the LP controller is remarkable.  On the 6.5 m observing a 0 mag star, the LP contrast is over 200 times better at 1 $\lambda/D$, and is nearly 10 times better for an 8th mag star. On the 25.4 m the LP contrast is over 1400 times better at 1 $\lambda/D$ on a 0 mag star, and is over 30 times better for an 8th mag star.

\afterpage{\clearpage}

\section{Discussion}
\label{sec:discussion}

\subsection{Predictive Control}
Predictive control has been discussed in many other studies.  Here we used a relatively simple form of it, rather than more sophisticated options.  The LP controller provides a conceptually simple extension of the integrator controller, and provides an efficient and easy to understand method for calculating its coefficients since we know the open-loop PSD.  However, it is not the most sophisticated method proposed to-date in the literature and our intent here is not to propose it as the best one.  

 Dessenne et al. \cite{1998ApOpt..37.4623D,1999OptL...24..339D} presented a method using a linear filter, essentially identical to the technique we used.  Their method for optimizing the parameters was based on an explicit least squares minimization using past data, rather than the PSD, and allowed the coefficients $a$ and $b$ to differ.  This technique was demonstrated on sky.

We recently presented a related predictive control strategy, based on empirical orthogonal functions (EOFs)\cite{guyon_males}. EOFs are essentially a time-and-space Karhunen-Lo\`{e}ve basis, and this method allows for including non-WFS measurements, such as accelerometers.  Filter optimization is done on past time-domain data. We found essentially the same level of improvement in simulations.  

Johnson et al. \cite{2011JOSAA..28.1566J} developed a control strategy based on predicting the wind-driven translation of turbulent layers.  With simulations incorporating on-sky data from the ViLLaGEs AO system \cite{2010SPIE.7736E..1OM} they showed significant increase in Strehl on fainter stars, and predicted large gains in Strehl for larger telescopes.  

Much recent work has been in the area of linear quadratic Gaussian (LQG) controllers.   Poyneer et al.\cite{2007JOSAA..24.2645P} proposed a Fourier-mode by Fourier-mode optimization of a Kalman filter controller, and using end-to-end simulations showed large gains in contrast similar to what we have shown.  They presented their method as a control filter much as we did here.  They pointed out that any filter could be cast in the direct form as in Equation (\ref{eqn:H_con}), though it is then subject to errors due to numerical precision if actually used to implement the filter. 

More broadly, the determination of optimum control filters is of great interest in the AO-related literature.  In the context of AO this has been studied extensively for LQG controllers \cite{2008OExpr..16...87P, 2012JOSAA..29..185C}.  See also the review of various system identification strategies presented in Kulcs\'ar et al.\cite{2012SPIE.8447E..0ZK}

While this paper was under review, Correia et al. presented an analysis of predictive control using the distributed Kalman filter method\cite{2017JOSAA..34.1877C}.  Their analysis employed the spatial filtering method (as opposed to our temporal frequency method), and showed post-coronagraph contrast gains for the Keck AO system at least as large as we show here.  

Any of these control methods can be represented by their temporal transfer functions, and so could be included in further studies using our technique.  In summary, the LP is not the only, nor the most advanced, technique we have available, but it is apparent that predictive control offers significant gains for high contrast imaging of exoplanets.

% The method we have just presented is more suitable for this analysis, since we have the open loop PSD {\it a priori}, which gives us the autocorrelation.  Thus we avoid having to generate a set of randomized time-series.  Though we are not here addressing the details of real-world implementation,  \cite{2012OExpr..2027108A} show that there are advantages in practice to using the PSD.  We note that existing modally optimized AO systems, including both SPHERE and GPI, estimate open-loop modal PSDs for use in gain optimization \cite{2014SPIE.9148E..0OP, 2016ApOpt..55..323P}. So, in principle, the highly efficient {\it Yule-Walker} technique for calculating the LP coefficients could be implemented on such systems.

\subsection{Sensitivity to Frozen Flow}

Our model of input turbulence was based on the frozen flow model.  However, our semi-analytic method does not require frozen flow.  That is, any source of the temporal PSDs could be used as input to the analysis.  We could incorporate boiling, or disturbances derived from telescope models (e.g. vibrations due to wind shake).  This could also be extended to assess the impact of time-variable turbulence profiles.

It should be expected that inclusion of additional sources of error will reduce projected performance.  In the case of, e.g., telescope vibrations this is simply because there is more variance in the system.  In the case of time-variable turbulence, i.e. changing wind and seeing, the reduced performance will result from sub-optimality of the filter coefficients.

% \subsection{Incorporating Other Terms}
% Here we dealt only with the time-lag \& photon-noise limited contrast, which is generally the dominant error.  The temporal PSD of other  terms can be calculated using the same recipe.  Once the residual is determined in the WFS wavelength, it can be scaled to arbitrary science wavelengths.  Using phase as an example, the first step is to divide by the $X(\lambda_{wfs})$ function, then scale by $1/\lambda^2$, then multiply by the $X(\lambda_{sci})$ function.  Errors corresponding to the chromaticity of scintillation and the index of refraction can then be included as well.

\subsection{Impact on Exoplanet Characterization}

Our motivation for this study was to analyze the limits of ground-based coronagraphic exoplanet characterization, and to assess the impact that predictive control will have on it.  Consider the coupling of an ExAO-fed coronagraph to a high-dispersion spectrograph\cite{2015A&A...576A..59S}, a technique recently dubbed  ``high dispersion coronagraphy'' (HDC) \cite{2017AJ....153..183W, 2017ApJ...838...92M}.  We focus on this technique because it is, in principle, photon-noise limited, meaning we do not have to consider speckle noise.  Neglecting detector and background noise, the signal-to-noise ratio (S/N) in the HDC technique is given by \cite{2015A&A...576A..59S}
\begin{equation}
\mbox{S/N} = \frac{ F_p \Delta t}{ \sqrt{F_* \Delta t \left<C\right>}}\sqrt{N_{\mbox{lines}}}.
\end{equation}
Where $F_p$ is the photon rate from the planet and $F_*$ is the photon rate from the star (both in one spectral resolution element),  and  $N_{\mbox{lines}}$ is the number of resolved lines in the planet spectrum used for cross-correlation template matching. $\left<C\right>$ is the raw PSF contrast as we have derived here. $\Delta t$ is the integration time needed to reach a desired S/N.    This leads directly to
\begin{equation}
\Delta t \propto \left<C\right>.
\end{equation}
So, in terms of required integration times the potential benefits of predictive control are quite profound.  Using just the relatively simple form of it considered in this study, we find that predictive control of atmospheric turbulence could make a GSMT up to 30 times more efficient at exoplanet characterization around a star as bright as Proxima Centauri, and up to 1400 times more efficient on brighter stars.

\subsection{Code Availability}

The code we developed for this analysis is freely available at \url{https://github.com/jaredmales/aoSystem}.  It is written in \verb c++ , is highly parallelized, and compiles to a command line application which accepts configuration files to set up the analysis.  It depends on the library available at \url{https://github.com/jaredmales/mxlib}, as well as several easily available open-source libraries.  No licenses are needed.

\section{Conclusion}
\label{sec:conclusion}

We have developed a semi-analytic framework for predicting the raw PSF contrast in an AO-fed coronagraphic image under closed-loop control.  Instead of the usual spatial-filter analysis, we explicitly calculated temporal PSDs and analyzed AO performance in the temporal-frequency domain.  Our technique takes into account closed-loop dynamics as well as a multi-layer frozen-flow turbulence model.  We also analyzed performance of ideal coronagraphs, showing how to account for the subtle difference between the residual PSD and the residual modal variance on an aperture.  Using our model, we showed that predictive control will improve raw contrast by up to a factor of 1400 on bright stars, and by a factor of 30 on 8th magnitude stars, compared to the simple integrator control law.  Assuming a photon-noise limited observation, the exposure time required to characterize an exoplanet will decrease by the same large factors.  This has significant implications for the potential of the coming ground-based GSMTs to search for life outside our Solar System, and makes predictive control one of the key techniques that must be perfected for exoplanet characterization.

\section{Acknowledgements}

This work was motivated by a question from Marcos van Dam. We thank Richard Frazin for reviewing an early draft of this manuscript.  We thank the two anonymous referees for their careful and insightful reviews which significantly improved this manuscript.  Early steps in this work were performed under contract with the California Institute of Technology (Caltech)/Jet Propulsion Laboratory (JPL) funded by NASA through the Sagan Fellowship Program executed by the NASA Exoplanet Science Institute.
The authors acknowledge support from NSF awards 1506818 and 1625441.

\clearpage
\appendix
\numberwithin{equation}{section}

\section{Mathematics of the Fourier Basis}
We derive the normalizations and Fourier transforms of the Fourier basis on a circular aperture.  
\subsection{Normalization}
\label{app:norm}
The $L^2$ norm of the cosine mode on the aperture is
\begin{equation}
\left|\cos(2\pi \vec{k}_{mn}\cdot \vec{q}) \right|^2 = \int \mathscr{A}(\vec{q})\cos^2(2\pi \vec{k}_{mn}\cdot \vec{q}) d\vec{q}
\end{equation}
For a symmetric aperture, we can restrict ourselves to $n=0$ with no loss of generality, which simplifies the integral to
\begin{equation}
\left|\cos(2\pi \vec{k}_{mn}\cdot \vec{q}) \right|^2 = \frac{4}{\pi D^2} \int_0^{D/2} \int_0^{2\pi} \cos^2(2\pi \textstyle\frac{m}{D} q \cos \theta) d\theta q dq
\end{equation}
Using the exponential form of the cosine
\begin{equation}
\left|\cos(2\pi \vec{k}_{mn}\cdot \vec{q}) \right|^2 = \int_0^{D/2} \int_0^{2\pi} \frac{1}{2} + \frac{1}{4}\left[ e^ {4\pi \textstyle\frac{m}{D} q \cos \theta} + e^ {-4\pi \textstyle\frac{m}{D} q \cos \theta} \right] d\theta q dq
\end{equation}
With the following identity
\begin{equation}
2\pi J_0(z) = \int_0^{2\pi} \mathrm{e}^{\pm i z \cos(\phi)} d\phi
\end{equation}
 (cf. \cite{2007tisp.book.....G} 8.411 \#7), the integral evaluates as
\begin{equation}
\left|\cos(2\pi \vec{k}_{mn}\cdot \vec{q}) \right|^2 = \frac{1}{2} + \frac{4}{D^2} \int_{0}^{D/2} J_0(4\pi \textstyle\frac{m}{D} q) qdq
\end{equation}
Next using the identity  
\begin{equation}
z J_1(z) = \int z J_0(z) dz
\end{equation}
(cf. \cite{2007tisp.book.....G} 5.52 \#1) and a change of variables we have
\begin{equation}
\left|\cos(2\pi \vec{k}_{mn}\cdot \vec{q}) \right|^2 = \frac{1}{2} + \Ji(2\pi k_{mn} D)
\end{equation}
Similarly, the $L^2$ norm of the sine mode on the aperture is
\begin{equation}
\left|\sin(2\pi \vec{k}_{mn}\cdot \vec{q}) \right|^2 = \frac{1}{2} - \Ji(2\pi k_{mn} D) 
\end{equation}
and the norm of the product is
\begin{equation}
\left|\sin(2\pi \vec{k}_{mn}\cdot \vec{q}) \cos(2\pi \vec{k}_{mn}\cdot \vec{q}) \right| = 0
\end{equation}

%----------------------------------------------------------
%Fourier Transforms of Fourier Basis.
%----------------------------------------------------------
\subsection{ Fourier transforms on an aperture}
\label{app:ft}
We need the Fourier transform of the Fourier modal basis over the aperture of the telescope, that is
\begin{equation}
{Q}_{mn}^c(\vec{k}) = \int_{-\infty}^{\infty} \int_{-\infty}^{\infty} \mathscr{A}(u,v)\cos(2\pi \vec{k}_{mn}\cdot \vec{q}) \mathrm{e}^{i2\pi(k_uu+k_vv)} \mathrm{d}u \mathrm{d}v
\end{equation}
The unobstructed circular aperture is described by Equation (\ref{eqn:unob_ap}).

Now switching to polar coordinates and collecting terms we have
\begin{equation}
{Q}_{mn}^c(\vec{k}) = \frac{4}{2\pi D^2}\int_{0}^{D/2} \int_{0}^{2\pi} \left(  \mathrm{e}^{i2\pi q\left[ \left(k_u+\frac{m}{D}\right)\cos\theta + \left(k_v+\frac{n}{D}\right)\sin\theta \right]} + \mathrm{e}^{-i2\pi q\left[ \left(k_u-\frac{m}{D}\right)\cos\theta + \left(k_v-\frac{n}{D}\right)\sin\theta \right]} \right) \mathrm{d}\theta q \mathrm{d}q
\end{equation}
where we have made use of the complex exponential form of the cosine.  We next introduce the notational simplification
\begin{align}
\begin{split}
k_{mn}^+ &= \sqrt{ \left(k_u + \frac{m}{D} \right)^2 + \left(k_v + \frac{n}{D} \right)^2 }\\
k_{mn}^- &= \sqrt{ \left(k_u - \frac{m}{D} \right)^2 + \left(k_v - \frac{n}{D} \right)^2 }
\end{split}
\end{align}
and the associated angles defined by
\begin{align}
\begin{split}
\cos \alpha = \left(k_u + \frac{m}{D} \right)/k_{mn}^+ & \qquad \sin \alpha = \left(k_v + \frac{n}{D} \right)/k_{mn}^+\\
\cos \beta = \left(k_u - \frac{m}{D} \right)/k_{mn}^- & \qquad \sin \beta = \left(k_v - \frac{n}{D} \right)/k_{mn}^-
\end{split}
\end{align}
The transform can now be written
\begin{equation}
{Q}_{mn}^c(\vec{k}) = \frac{2}{ \pi D^2}\int_{0}^{D/2} \int_{0}^{2\pi} \left(  \mathrm{e}^{i2\pi q k_{mn}^+\cos(\theta - \alpha) } + \mathrm{e}^{-i2\pi q k_{mn}^- \cos (\theta-\beta) } \right) \mathrm{d}\theta q \mathrm{d}q
\end{equation}
With the following identity
\begin{equation}
2\pi J_0(z) = \int_0^{2\pi} \mathrm{e}^{\pm i z \cos(\phi)} d\phi
\end{equation}
 (cf. \cite{2007tisp.book.....G} 8.411 \#7)  the transform is equivalent to
\begin{equation}
{Q}_{mn}^c(\vec{k}) = \frac{4}{D^2} \int_{0}^{D/2} \left( J_0(2\pi k_{mn}^+ q) + J_0(2\pi k_{mn}^- q)\right) q \mathrm{d}q
\end{equation}
where $J_n(x)$ are the cylindrical Bessel functions of the first kind.  With the identity  
\begin{equation}
z J_1(z) = \int z J_0(z) dz
\end{equation}
(cf. \cite{2007tisp.book.....G} 5.52 \#1) we finally have
\begin{equation}
{Q}_{mn}^c (\vec{k}) = \left[ \frac{ J_1( \pi D k_{mn}^+)}{\pi D k_{mn}^+} + \frac{ J_1( \pi D k_{mn}^-)}{\pi D k_{mn}^-} \right].
\end{equation}
Similarly for the sine modes 
\begin{equation}
{Q}_{mn}^s(\vec{k}) = i \left[ \frac{ J_1( \pi D k_{mn}^+)}{\pi D k_{mn}^+} - \frac{ J_1( \pi D k_{mn}^-)}{\pi D k_{mn}^-} \right].
\end{equation}

\section{Contrast $C_2$ of Guyon (2005)}
\label{app:guyon_contrast}

Here we present a derivation of the contrast due to time delay and measurement error from \cite{2005ApJ...629..592G}, there called $C_2$.  We extend the analysis of \cite{2005ApJ...629..592G} to include finite loop delays, detector and background noise, and make it consistent with the results of Section \ref{sec:psf_contrast}.   A spatial frequency component of the wavefront phase at time $t$ can be written as
\begin{equation}
\Phi_{mn}(\vec{q},t) = \frac{2\pi}{\lambda} h_{mn}^\dagger(t) \cos \left( \vec{k}_{mn} \cdot \vec{q} + \phi_{mn}^\dagger(t) \right)
\end{equation}
where the coefficient $h_{mn}^\dagger(t)$ is the combined amplitude of the Fourier mode, and $\phi_{mn}^\dagger(t)$ is its phase.  The $\dagger$ superscript is to differentiate this unnormalized basis from the normalized Fourier modes presented in Section \ref{sec:basic_fourier}.  Now at a time $\tau_{tl}$ later the wavefront will be
\begin{equation}
\Phi_{mn}(\vec{q}, t+\tau_{tl}) \approx \Phi_{mn}(\vec{q}, t) + \frac{\partial \Phi_{mn}(\vec{q}, t)}{\partial t} \tau_{tl}
\end{equation}
The residual variance after applying the wavefront correction will be
\begin{equation}
\sigma_{tl,mn}^2 \approx \left< \left( \frac{\partial \Phi_{mn}(\vec{q}, t)}{\partial t}  \right)^2 \right>\: \tau_{tl}^2
\end{equation}
which becomes
\begin{equation}
\sigma_{tl,mn}^2 \approx \left(\frac{2 \pi}{\lambda} \right)^2 \left(\left<\left( \frac{dh_{mn}^\dagger(t)}{dt} \right)^2\right> + \left<\left( h_{mn}^\dagger(t)\frac{d\phi_{mn}^\dagger(t)}{dt}  \right)^2\right> \right) \tau_{tl}^2
\end{equation}

We next employ the frozen flow hypothesis, assuming that the amplitude does not change and that the phase changes only due to wind-driven flow.  This gives
\begin{align}
\begin{split}
\frac{dh_{mn}^\dagger(t)}{dt} &= 0 \\
\frac{d\phi_{mn}^\dagger(t)}{dt} &= 2\pi \vec{k}_{mn}\cdot\vec{\mathpzc{V}}
\end{split}
\end{align}
which leads to the residual variance due to time lag
\begin{equation}
\sigma_{tl,mn}^{2\;\dagger} \approx \left(\frac{2 \pi}{\lambda} \right)^2 \left< h_{mn}^{\dagger\;2}(t)\right> \left(2\pi   \vec{k}_{mn}\cdot\vec{\mathpzc{V}} \right)^2 \tau_{tl}^2.
\end{equation}

Now the variance of each mode will be given by the spatial PSD according to
\begin{equation}
\left< h_{mn}^{\dagger\;2}(t)\right> = \frac{ \mathcal{P}(\vec{k}_{mn})}{D^2} \left(\frac{\lambda_0}{2\pi}\right)^2 \mbox{ [m}^2\mbox{ rms]}.
\end{equation}

Employing the wavefront variance from measurement noise given by Equation (\ref{eqn:sigma_ph_f_sn}), the total residual variance at the science wavelength will be
\begin{equation}
\sigma_{mn}^{\dagger\;2} = \left(\frac{\lambda_0}{\lambda} \right)^2 \frac{\mathcal{P}(\vec{k})}{D^2} \left(2\pi   \vec{k}_{mn}\cdot\vec{\mathpzc{V}} \right)^2 \tau_{tl}^2 + \frac{ \beta_p^2(\vec{k}_{mn}) }{ S/N^2} \left(\frac{\lambda_{wfs}}{\lambda}\right)^2 \mbox{ [rad}^2\mbox{ rms.]}
\label{eqn:c2_totvar}
\end{equation}
In this open-loop framework, the equivalent time lag is $\tau_{tl} = \tau_{wfs} + \tau$ where $\tau$ is the closed loop delay defined in Section \ref{sec:xfer_funcs}, and $\tau_{wfs}$ accounts for the sample-and-hold of the WFS and DM  \cite{1999aoa..book.....R_ch6}.  We calculate the optimum frame rate which minimizes the variance by differentiating Equation (\ref{eqn:c2_totvar}) with respect to $\tau_{wfs}$, which yields a quartic equation
\begin{equation}
\lambda_{0}^2\frac{\mathcal{P}(\vec{k}_{mn})}{D^2} \left(2\pi   \vec{k}_{mn}\cdot\vec{\mathpzc{V}} \right)^2\left[\tau_{wfs}^4 +
 \tau \tau_{wfs}^3\right] -
\frac{\lambda_{wfs}^2 \beta_p^2(\vec{k}_{mn})}{F_\gamma^2} \left[    (F_\gamma + n_{px}F_{bg})\tau_{wfs} + 2n_{px}\sigma_{ron}^2\right] = 0.
\end{equation}
Each mode will have its own optimum $\tau_{wfs}$, which is somewhat analogous to modal gains in closed-loop control.  Since $\left<h_{mn}^{p\;2}\right> = \left<h_{mn}^{c\dagger\;2}\right> + \left<h_{mn}^{s\dagger\;2}\right> = \left<h_{mn}^{\dagger\;2}\right>$, we can use the results of Section \ref{sec:psf_contrast} to write
\begin{equation}
\left< I_{\Phi, mn}^\dagger(\vec{r}) \right> = \sigma_{mn}^{\dagger\;2} \left[ \mbox{PSF}(\vec{r}-\vec{k}_{mn}\lambda) +  \mbox{PSF}(\vec{r}+\vec{k}_{mn}\lambda)\right].
\label{eqn:contrast_var_mn_c2}
\end{equation}
The summation in Equation (\ref{eqn:contrast_Phi}) defines $\left< I_{\Phi}^\dagger( r ) \right>$, which leads to the expression for the contrast $C_2$
\begin{equation}
C_2(\vec{r})   = \frac{ \left< I_\Phi^\dagger( \vec{r} ) \right>  }{ \left< S \right> \: \mathrm{PSF}(0)}.
\label{eqn:contrast_c2}
\end{equation}
The Strehl ratio can be calculated by summing the total WFE after the convolution
% \begin{equation}
% \sigma^2 = \sum_{mn}
% \begin{cases}
% \sigma_{mn}^{\dagger\;2} & \mbox{if } |m| < \frac{D}{2d}, |n| < \frac{D}{2d} \\
% \frac{ \mathcal{P}(\vec{k}_{mn})}{D^2} \left(\frac{\lambda_0}{\lambda}\right)^2 & \mbox{ otherwise}
% \end{cases}
% \end{equation}
and then employing the Marechal approximation
\begin{equation}
\left< S \right> = e^{-\sigma^2}
\end{equation}
It is straightforward to derive the other contrasts, $C_{0,1,3,4,5,6}$, of \cite{2005ApJ...629..592G} using similar arguments.

\section{Calculating the Autocorrelation}
\label{app:autocorr}
The autocorrelation $\mathcal{R}$ can be calculated from a numerical PSD using the following recipe:

1. Start with temporal PSD $\mathcal{T}_{mn} (f)$ at $N$ discrete frequencies $f_j$, where $f_1 = \Delta f$ and $f_N = \frac{1}{2} f_s$.

2. Form the two-sided temporal PSD of length $2N$ by augmenting with 0 at the beginning, and augmenting with the reverse at the end:
\begin{equation}
\mathcal{T}_{mn}^{'} (f_{j'}) =
\begin{cases}
 0, & j' = 1\\
\mathcal{T}_{mn} (f_j) , & j' = 2 \dots N+1, j=1\dots N\\
\mathcal{T}_{mn} (f_j) , & j' = N+2 \dots 2N, j=N-1\dots 1
\end{cases}
\end{equation}

3. Calculate the DFT of the augmented PSD using the FFT algorithm, giving the autocorrelation at lag $\tau$
\begin{equation}
\mathcal{R}_{mn}(\mathpzc{t}) = \mbox{FFT}\left\{\mathcal{T}_{mn}^{'} (f_{j'}) \right\}
\end{equation}
Note that this is will be a circular autocorrelation wrapping around after $N$ points. $N$ should be chosen to be larger (by at least a factor of 2) than the needed number of points of $\mathcal{R}$.

4.  The lags of the points are given by
\begin{equation}
\mathpzc{t}_{j'} = 
\begin{cases}
\frac{j'-1}{2Nf_s}, & j'=1 \dots N+1 \\
\frac{j'-2N-1}{2Nf_s}, & j'= N+2 \dots 2N
\end{cases}
\end{equation}
where only the first $N+1$ points are unique.
\clearpage
\bibliographystyle{spiejour}
\footnotesize
\bibliography{males_ao_closed_loop}

\end{document}